\documentclass[aps,floats]{revtex4}
\usepackage{amsmath,amssymb}
\usepackage{graphicx,epsfig}
\usepackage[greek,english]{babel}
\usepackage{bbold}

\usepackage{hyperref}

\begin{document}
\bibliographystyle {plain}
\pdfoutput=1
\def\oppropto{\mathop{\propto}} 
\def\opsimeq{\mathop{\simeq}}
\def\opoverderline{\mathop{\overline}}
\def\operarrow{\mathop{\longrightarrow}}
\def\opsim{\mathop{\sim}}

\def\opmin{\mathop{\min}} 
\def\opmax{\mathop{\max}} 
\def\oplim{\mathop{\lim}}

\def\fig#1#2{\includegraphics[height=#1]{#2}}
\def\figx#1#2{\includegraphics[width=#1]{#2}}


\title{ Revisiting the Ruelle thermodynamic formalism for Markov trajectories \\
with application to the glassy phase of random trap models } 


\author{ C\'ecile Monthus }
 \affiliation{Institut de Physique Th\'{e}orique, 
Universit\'e Paris Saclay, CNRS, CEA,
91191 Gif-sur-Yvette, France}

\begin{abstract}
The Ruelle thermodynamic formalism for dynamical trajectories over the large time $T$ corresponds to the large deviation theory for the information per unit time of the trajectories probabilities. The microcanonical analysis consists in evaluating the exponential growth in $T$ of the number of trajectories with a given information per unit time, while the canonical analysis amounts to analyze the appropriate non-conserved $\beta$-deformed dynamics in order to obtain the scaled cumulant generating function of the information, the first cumulant being the famous Kolmogorov-Sinai entropy. This framework is described in detail for discrete-time Markov chains and for continuous-time Markov jump processes converging towards some steady-state, where one can also construct the Doob generator of the associated $\beta$-conditioned process. The application to the Directed Random Trap model on a ring of $L$ sites allows to illustrate this general framework via explicit results for all the introduced notions. In particular, the glassy phase is characterized by anomalous scaling laws with the size $L$ and by non-self-averaging properties of the Kolmogorov-Sinai entropy and of the higher cumulants of the trajectory information.

\end{abstract}

\maketitle

\section{ Introduction }

Ruelle thermodynamic formalism for dynamical trajectories 
has allowed to make the link between the field of dynamical chaotic systems and the statistical physics of equilibrium \cite{ruelle} and non-equilibrium \cite{beck,gaspard_review,dorfman}.
In particular, the application of Ruelle thermodynamic formalism to Markov processes \cite{gaspard_review,gaspard2004,vivien_thesis,lecomte_chaotic,lecomte_thermo,lecomte_formalism,lecomte_glass,kristina1,kristina2,c_dynentropy} has put forward the Kolmogorov-Sinai entropy as an essential observable to characterize the stochastic trajectories.
The unifying language is actually the theory of large deviations 
(see the reviews \cite{oono,review_touchette} and references therein) that has more generally lead to major advances in the analysis of
non-equilibrium steady-states (see the reviews with different scopes \cite{derrida-lecture,harris_Schu,searles,harris,mft,sollich_review,lazarescu_companion,lazarescu_generic,jack_review}, the PhD Theses \cite{fortelle_thesis,vivien_thesis,chetrite_thesis,wynants_thesis} 
 and the HDR Thesis \cite{chetrite_HDR}). 
In particular, the analysis of generating functions of time-additive observables via deformed Markov operators
has been used extensively in many models  \cite{derrida-lecture,sollich_review,lazarescu_companion,lazarescu_generic,jack_review,vivien_thesis,lecomte_chaotic,lecomte_thermo,lecomte_formalism,lecomte_glass,kristina1,kristina2,jack_ensemble,simon1,simon2,simon3,Gunter1,Gunter2,Gunter3,Gunter4,chetrite_canonical,chetrite_conditioned,chetrite_optimal,chetrite_HDR,touchette_circle,touchette_langevin,touchette_occ,touchette_occupation,derrida-conditioned,derrida-ring,bertin-conditioned,touchette-reflected,touchette-reflectedbis,c_reset,c_lyapunov,previousquantum2.5doob,quantum2.5doob,quantum2.5dooblong},
 with the formulation of the corresponding 'conditioned' process via the generalization of Doob's h-transform.

Within this perspective, the Ruelle thermodynamic formalism can be rephrased as the large deviation theory of the information per unit time of the trajectories probabilities. For Markov processes, this information is a time-additive observable, whose average is the Kolmogorov-Sinai entropy, while it is also important to analyze its higher cumulants, in particular its variance. It is thus interesting to revisit the Ruelle thermodynamic formalism both for discrete-time Markov chains and for continuous-time Markov Jump processes in order to study the generating function of the information via the appropriate $\beta$-deformed Markov generator
and the Doob generator of the corresponding $\beta$-conditioned process. 
As examples of applications where this analysis leads to very explicit results, 
we will consider the Directed Random Trap Model 
\cite{jpdir,aslangul,comptejpb,directedtrapandsinai},
whose large deviations properties have been previously studied for the current \cite{vanwijland_trap,c_ring}
and from the point of view of inference \cite{c_inference}.
Note that besides this Directed Random Trap Model, many other trap models
have been also analyzed in the context of anomalously slow glassy behaviors 
\cite{jpb_weak,dean,jp_ac,jp_pheno,bertinjp1,bertinjp2,trapsymmetric,trapnonlinear,trapreponse,trap_traj}.
Our main conclusion will be that the glassy phase of the Directed Random Trap Model on a ring of size $L$ can be characterized via the anomalous scaling with the size $L$ and the non-self-averaging properties of the Kolmogorov-Sinai entropy and of the higher cumulants of the information.

The paper is organized as follows.
In section \ref{sec_ruelle}, the Ruelle thermodynamic formalism for dynamical trajectories is described as the large deviation theory for the information per unit time of the trajectories probabilities, with the corresponding microcanonical and canonical perspectives.
The application to discrete-time Markov Chains with steady-state is described in \ref{sec_chain}, 
while the example of the discrete-time random trap model on the ring is studied in section \ref{sec_chaintrap}.
The application to continuous-time Markov Jump processes with steady-state is presented in \ref{sec_jump}, 
while the example of the continuous-time random trap model on the ring is analyzed in section \ref{sec_jumptrap}.
Our conclusions are summarized in section \ref{sec_conclusion}. Appendix \ref{app_per} contains a reminder on the perturbation theory for an isolated eigenvalue of non-symmetric matrix, while Appendix \ref{app_levy} contains a reminder on L\'evy stable laws.


\section{ Reminder on the Ruelle thermodynamic formalism for trajectories }

\label{sec_ruelle}

One considers all the possible stochastic trajectories $x(0 \leq t \leq T)$ over the large time $T$
 with their probabilities 
or probability densities ${\cal P}[x(0 \leq t \leq T)]  $
normalized to unity
\begin{eqnarray}
1= \sum_{x(0 \leq t \leq T)}  {\cal P}[x(0 \leq t \leq T)] \equiv  \sum_{x(.)}  {\cal P}[x(.)]
\label{normaptraj}
\end{eqnarray}
where the last simplified notations will be used in order to ease the read of some equations.
The average of an observable $A[x(0 \leq t \leq T)]$ of the trajectory $x(0 \leq t \leq T)$
with respect to the probability measure of Eq. \ref{normaptraj} will be denoted by
\begin{eqnarray}
< A[x(.)]> \equiv  \sum_{x(.)}  {\cal P}[x(.)] A[x(.)]
\label{averaptraj}
\end{eqnarray}

\subsection{ Microcanonical analysis : counting the trajectories with a given intensive information $I$ }

In the microcanonical analysis, the trajectories $x(0 \leq t \leq T)$ 
are classified with respect to their intensive information $I $, i.e. their information per unit time
\begin{eqnarray}
 I[x(0 \leq t \leq T)] \equiv - \frac{ \ln   {\cal P}[x(0 \leq t \leq T)] }{T}
\label{infotraj}
\end{eqnarray}
which is a positive observable $I[x(.)] \geq 0$ as a consequence of the normalization of Eq. \ref{normaptraj}.

The number of trajectories with a given intensive information $I$
\begin{eqnarray}
\Omega_T(I) \equiv \sum_{x(0 \leq t \leq T)}    \delta \left( I - I[x(0 \leq t \leq T)] \right)
\label{numberi}
\end{eqnarray}
 grows exponentially with the time $T$
\begin{eqnarray}
\Omega_T(I) \opsimeq_{T \to +\infty}  e^{T s(I) }
\label{numberigrowth}
\end{eqnarray}
where $s(I) $ represents the intensive microcanonical Boltzmann entropy of trajectories with intensive information $I$.
The region of positive entropy $s(I) \geq 0$ corresponding 
to numbers $\Omega_T(I) \geq 1 $ of trajectories
 defines the interval $[I^{min},I^{max}]$ of possible information $I$
\begin{eqnarray}
s( I ) \geq 0 \ \ \ \ {\rm for } \ \ \ \ \ I^{min} \leq I \leq I^{max}
\label{intervalinfo}
\end{eqnarray}
The physical meaning is that
$I^{min}$ corresponds to the trajectory with the highest individual trajectory probability 
\begin{eqnarray}
 \max_{x(0 \leq t \leq T)}  {\cal P}[x(0 \leq t \leq T)] \opsimeq_{T \to +\infty} e^{-T I^{min} }
\label{maxptraj}
\end{eqnarray}
while $I_{max}$ corresponds to the trajectory with the smallest individual trajectory probability 
\begin{eqnarray}
 \min_{x(0 \leq t \leq T)}  {\cal P}[x(0 \leq t \leq T)] \opsimeq_{T \to +\infty} e^{-T I^{max} }
\label{minptraj}
\end{eqnarray}


\subsection{ Canonical analysis via the dynamical partition function in terms of the parameter $\beta$ }

The dynamical partition function defined in terms of the parameter $\beta $ 
\begin{eqnarray}
Z_T(\beta) \equiv \sum_{x(0 \leq t \leq T)}  \left( {\cal P}[x(0 \leq t \leq T)] \right)^{\beta}
\equiv  \sum_{x(.)} \left( {\cal P}[x(.)] \right)^{\beta}
\label{zbeta}
\end{eqnarray}
displays the exponential dependence  with the time $T$
\begin{eqnarray}
Z_T(\beta) \opsimeq_{T \to +\infty}  e^{T \psi(\beta) }
\label{zbetagrowth}
\end{eqnarray}
where the sign of the coefficient $\psi(\beta)$ changes at $ \psi(\beta=1)=0$ 
as a consequence of the normalization of Eq. \ref{normaptraj}.
\begin{eqnarray}
 \psi(\beta) \geq 0 \ \ \ \ {\rm for } \ \ \ \ \ \beta \leq 1
 \nonumber \\
  \psi(\beta) \leq 0 \ \ \ \ {\rm for } \ \ \ \ \ \beta \geq 1
\label{signepsi}
\end{eqnarray}

Via the change of variables of Eq. \ref{infotraj},
the dynamical partition function of Eq. \ref{zbeta}
can be rewritten as an integral over the intensive information $I$ using Eqs \ref{numberi} , \ref{numberigrowth}
and \ref{intervalinfo}
\begin{eqnarray}
Z_T(\beta) = \int_{I^{min}}^{I^{max}} dI \Omega_T(I)   e^{-\beta T I} \opsimeq_{T \to +\infty}  
\int_{I^{min}}^{I^{max}} dI    e^{ T \left[ s(I) -\beta  I \right]} 
\label{zbetacol}
\end{eqnarray}
For large time $T\to +\infty$, 
the saddle-point evaluation of this integral yields that the function $ \psi(\beta)$
introduced in Eq. \ref{zbetagrowth}
corresponds to the Legendre transform of the microcanonical Boltzmann entropy $s(I)$
\begin{eqnarray}
s'(I) -\beta  && =0
\nonumber \\
s(I) -\beta  I && = \psi(\beta)
\label{legendre}
\end{eqnarray}
while the reciprocal Legendre transform reads
\begin{eqnarray}
0  && =\psi'(\beta) +  I
\nonumber \\
s(I)  && = \psi(\beta) + \beta  I
\label{legendrereci}
\end{eqnarray}
This Legendre transform holds in the interval $I^{min} \leq I \leq I^{max}$ of Eq. \ref{intervalinfo},
i.e. for $\beta_{I^{max}} \leq \beta \leq  \beta_{I^{min}}$ where
\begin{eqnarray}
\beta_{I^{min}}  && =s'(I^{min})
\nonumber \\
\beta_{I^{max}}  && =s'(I^{max})
\label{legendrebord}
\end{eqnarray}
Outside the interval $\beta_{I^{max}} \leq \beta \leq  \beta_{I^{min}}$, 
the saddle-point will remain frozen at the boundary $I^{min}$ for $\beta > \beta_{I^{min}} $
and at the boundary $I^{max}$ for $\beta < \beta_{I^{max}} $ with the corresponding linear behaviors
\begin{eqnarray}
 \psi(\beta)  && = - \beta I_{min}  \ \ \ \ {\rm for } \ \ \ \ \ \beta > \beta_{I^{min}} 
 \nonumber \\
  \psi(\beta)  && = - \beta I_{max}  \ \ \ \ {\rm for } \ \ \ \ \ \beta < \beta_{I^{max}} 
\label{zbetainfty}
\end{eqnarray}
Let us now discuss some important values of the parameter $\beta$.


\subsection{ The special value $\beta=0$ and the corresponding information $I_{\beta=0}$ }

For $\beta=0$, the dynamical partition function of Eq. \ref{zbeta}
\begin{eqnarray}
Z_T(\beta=0)  = \sum_{x(0 \leq t \leq T)} 1 \opsimeq_{T \to +\infty}  
\int_{I^{min}}^{I^{max}} dI    e^{ T  s(I) } 
  \opsimeq_{T \to +\infty}  e^{T \psi(0) }
\label{zbetazero}
\end{eqnarray}
simply counts the total number of possible dynamical trajectories of length $T$.
The value $\beta=0 $ 
is associated via the Legendre transform of Eq. \ref{legendre}
to the value $I_{\beta=0}$ that maximises the microcanonical entropy $s(I)$
\begin{eqnarray}
s'(I_0)   && =0
\nonumber \\
s(I_0)  && = \psi(\beta=0)
\label{legendrezero}
\end{eqnarray}
i.e. the whole set of trajectories is dominated by the subset of trajectories having exactly this information $I_0$.


\subsection{ Series expansion of $\psi(\beta)$ around $\beta=1$ to obtain the cumulants of the intensive information}

Via the change of notation $\beta=1+\epsilon$, the dynamical partition function of Eq. \ref{zbeta} can be rewritten as
the generating function of the moments of the information of Eq. \ref{infotraj}
\begin{eqnarray}
Z_T(\beta=1+\epsilon)   && \equiv  \sum_{x(.)}  {\cal P}[x(.)] e^{ -\epsilon T I[x(.)] }  
=  \sum_{x(.)}  {\cal P}[x(.)] \left[ 1  -\epsilon T I[x(.)] + \frac{\epsilon^2 }{2} T^2  I^2[x(.)]  + \sum_{n=3}^{+\infty} \frac{(-\epsilon)^n }{n!} T^n  I^n[x(.)] \right]
\nonumber \\
&& 
=  1  -\epsilon T <I[x(.)] >+ \frac{\epsilon^2 }{2} T^2  <I^2[x(.)] > + \sum_{k=3}^{+\infty} \frac{(-\epsilon)^n }{k!} T^n < I^n[x(.)] >
\nonumber \\
&&   \opsimeq_{T \to +\infty}  e^{T \psi(1+\epsilon) }
=  e^{ \displaystyle T \left[ \epsilon \psi'(1) +\frac{\epsilon^2}{2} \psi''(1)
+ \sum_{n=3}^{+\infty}  \frac{\epsilon^n}{n!} \psi^{(n)}(1)\right] }
\label{zbeta1}
\end{eqnarray}
So $\psi(\beta=1+\epsilon)$ represents the scaled cumulant generating function 
of the information : its power expansion in $\epsilon$ around $\epsilon=0$ 
allows to evaluate the cumulant of order $n$ of the information $I[x(0 \leq t \leq T)]$
 in terms of the $n^{th}$ derivative $ \psi^{(n)}(1) $ of order $n$ at $\beta=1$.
Let us now discuss the physical meaning of the two first cumulants $n=1$ and $n=2$.

\subsubsection{ The averaged value of the information and the Kolmogorov-Sinai entropy $h_{KS}=-  \psi'(\beta=1)=I_{\beta=1}$ }

At order $\epsilon$, Eq. \ref{zbeta1} yields that the average of the intensive information $I[x(0 \leq t \leq T)]  $ over the trajectories $x(0 \leq t \leq T) $
converges for large time $T \to +\infty$ towards the opposite of the derivative of $\psi(\beta)  $ at $\beta=1$
\begin{eqnarray}
 <I[x(.)] > &&   \opsimeq_{T \to +\infty} -  \psi'(1) 
\label{infoav}
\end{eqnarray}
This averaged value is known as the Kolmogorov-Sinai entropy with the standard notation $h_{KS}$ 
\begin{eqnarray}
h_{KS} && \equiv 
\oplim_{T \to +\infty} \left( - \frac{1}{T}  \sum_{x(0 \leq t \leq T)}   {\cal P}[x(0 \leq t \leq T)] 
\ln \left( {\cal P}[x(0 \leq t \leq T)] \right)\right)
\nonumber \\
&& =  \oplim_{T \to +\infty} \left(  \sum_{x(.)}   {\cal P}[x(.)]  I[x(.)]  \right)  
\opsimeq_{T \to +\infty} -  \psi'(1)
\label{hksdef}
\end{eqnarray}
and describes the linear growth in $T$ of the 
Shannon entropy $ S^{dyn}(T)$ associated to the probability distribution of the dynamical trajectories
\begin{eqnarray}
S^{dyn}(T) \equiv  - \sum_{x(0 \leq t \leq T)}   {\cal P}[x(0 \leq t \leq T)] 
\ln \left( {\cal P}[x(0 \leq t \leq T)] \right) \oppropto_{T \to +\infty} T \ h_{KS}
\label{defdynentropy}
\end{eqnarray}
The reciprocal Legendre transform of Eq. \ref{legendrereci} for $\beta=1$ yields that the information value $I_{\beta=1}$
associated to $\beta=1$ 
\begin{eqnarray}
0  && =\psi'(1) +  I_1
\nonumber \\
s(I_1)  && =     I_1
\label{legendrerecibeta1}
\end{eqnarray}
coincides with the Kolmogorov-Sinai entropy 
\begin{eqnarray}
I_1   =- \psi'(1) = h_{KS}
\label{ibeta1}
\end{eqnarray}
that satisfies
\begin{eqnarray}
s(h_{KS})  && =     h_{KS}
\label{legendrereciks}
\end{eqnarray}
This means that the average over trajectories with their probabilities ${\cal P}[x(0 \leq t \leq T)] $
 is actually dominated by the number (see Eq. \ref{numberigrowth}) 
 \begin{eqnarray}
\Omega_T(h_{KS}) \opsimeq_{T \to +\infty}  e^{T s(h_{KS}) } = e^{T h_{KS} }
\label{numberigrowthks}
\end{eqnarray}
   of trajectories associated to the information value $I_1=h_{KS}$,
  where all these trajectories have the same probability given by  $ e^{-T h_{KS}}= \frac{1}{\Omega_T(h_{KS})}$.

\subsubsection{ The second derivative $\psi''(\beta=1)$ as the scaled variance of the intensive information }

The expansion at order in $\epsilon^2$ of Eq. \ref{zbeta1}
yields that the second derivative $\psi''(\beta=1)$ 
correspond to the scaled variance $V_{KS}$ of the intensive information $I[x(0 \leq t \leq T)]  $
  \begin{eqnarray}
V_{KS} &&  \equiv 
\oplim_{T \to +\infty} \left[  T 
\left( <I^2[x(.)]> - < I[x(.)]>^2 \right)
\right] \opsimeq_{T \to +\infty}  \psi''(1)
\label{psideri2}
\end{eqnarray}
  and thus characterizes the fluctuations around the averaged value $I_1= - \psi'(1) = h_{KS}$ discussed above.


\section{ Discrete-time Markov Chains with steady-state }

\label{sec_chain}

In this section, we focus on the Markov Chain dynamics for the probability $P_y(t)  $ to be at position $y$ at time $t$
\begin{eqnarray}
P_x(t+1) =  \sum_y W_{x,y}  P_y(t)
\label{markovchain}
\end{eqnarray}
with the normalization of the Markov matrix 
\begin{eqnarray}
  \sum_x W_{x,y}  && =1
\label{markovnorma}
\end{eqnarray}
in order to apply the formalism described in the previous section.

\subsection{ Steady-State and finite-time propagator }

We will assume that the normalized steady-state $P^*(x) $ of Eq. \ref{markovchain}
\begin{eqnarray}
P^*_x =  \sum_y W_{x,y}  P^*_y
\label{markovchainst}
\end{eqnarray}
exists. From the point of view of the Perron–Frobenius theorem,
Eqs \ref{markovnorma} and \ref{markovchainst} mean that unity is the highest eigenvalue of the positive Markov Matrix $W(.,.)$,
where the positive left eigenvector $l_x$ is trivial
\begin{eqnarray}
 l_x=1
\label{markovleft}
\end{eqnarray}
while  the right eigenvector $r(x)$ is the steady state
\begin{eqnarray}
 r_x=P^*_x
\label{markovright}
\end{eqnarray}
The whole spectral decomposition of the matrix $W$
\begin{eqnarray}
 W && = \vert r \rangle \langle l \vert + \sum_k e^{  -  \zeta_k } \vert \zeta^R_k \rangle \langle \zeta_k^L \vert 
\label{chainspectral}
\end{eqnarray}
involving the other eigenvalues $e^{- \zeta_k}<1$ labelled by the index $k$,
with their right eigenvectors $\vert \zeta^R_k \rangle $ and their left eigenvectors $ \langle \zeta_k^L \vert  $
satisfying the closure relation
\begin{eqnarray}
 \mathbb{1} && = \vert r \rangle \langle l \vert + \sum_k  \vert \zeta^R_k \rangle \langle\zeta_k^L  \vert 
\label{fermeture}
\end{eqnarray}
is useful to describe the relaxation 
of the finite-time propagator $\langle x \vert W^t \vert x_0 \rangle $ towards the steady state $P^*_x$
\begin{eqnarray}
\langle x \vert W^t \vert x_0 \rangle = 
\langle x \vert r \rangle \langle l \vert x_0 \rangle+ \sum_k e^{  -  t \zeta_k }\langle x \vert \zeta^R_k \rangle \langle \zeta_k^L \vert 
x_0 \rangle
= P^*_x + \sum_k e^{  - t \zeta_k } \langle x \vert \zeta^R_k \rangle \langle \zeta_k^L \vert x_0 \rangle
\label{chainpropa}
\end{eqnarray}


\subsection{ The intensive information $I$ as a time-additive observable of the trajectory $x(0 \leq t \leq T) $ }

When the initial position $x(0)$ is drawn with the steady-state distribution $P^*$ of Eq. \ref{markovchainst},
the probability of the trajectory $x(0 \leq t \leq T)$ reads
\begin{eqnarray}
{\cal P}[x(0 \leq t \leq T)]  =  \left[ \prod_{t=1}^T W_{x(t) , x(t-1)}\right]  P^*_{x(0)}
= \left[ \prod_{t=1}^T \langle x(t ) \vert W \vert x(t-1 ) \rangle \right] \langle x(0 ) \vert r \rangle
\label{pwtraj}
\end{eqnarray}
If one is interested only in the joint distribution of the positions $x(t_n) $ at the $N$ times $t_1 < t_2<..<t_N$,
one just needs to sum over the possible intermediate positions at the other times 
with the closure relation to obtain
\begin{eqnarray}
p [x(t_N);...x(t_2);x(t_1)]  = \left( \prod_{n=2}^{N-1}  \langle x(t_{n} ) \vert W^{(t_n-t_{n-1})} \vert x(t_{n-1} ) \rangle \right)  
 \langle x(t_1 ) \vert r \rangle
\label{pNpoints}
\end{eqnarray}
in terms of the finite-time propagator of Eq. \ref{chainpropa} and of the steady-state $ \langle x(t_1 ) \vert r \rangle
= P^*_{x(t_1)} $

The intensive information of Eq. \ref{infotraj} associated to the trajectories probabilities of Eq. \ref{pwtraj}
\begin{eqnarray}
I[x(0 \leq t \leq T)] \equiv - \frac{ \ln   {\cal P}[x(0 \leq t \leq T)] }{T}
=  - \frac{ 1 }{T} \sum_{t=1}^T \ln \left( W_{x(t) , x(t-1)} \right)  - \frac{ 1 }{T}\ln \left(  P^*_{x(0)}  \right)
\label{infotrajchain}
\end{eqnarray}
corresponds to the sum over the time $t$ of the time-local observable $\ln \left( W(x(t) , x(t-1)) \right) $
that characterizes the flows between two consecutive positions within the trajectory $x(0 \leq t \leq T) $.
As a consequence, the averaged value of the information with the trajectories probabilities of Eq. \ref{pwtraj}
only requires the knowledge of the partial distribution of Eq. \ref{pNpoints}
for $N=2$ consecutive times
\begin{eqnarray}
p [x(t);x(t-1)]  =  W_{x(t) , x(t-1)}  P^*_{x(t-1)}
\label{p2points}
\end{eqnarray}
in order to to compute
\begin{eqnarray}
< \ln \left( W_{x(t) , x(t-1)} \right) > 
&& = \sum_{x(t)} \sum_{x(t-1)} p [x(t);x(t-1)]  \ln \left( W_{x(t) , x(t-1)} \right)
=  \sum_{x(t)} \sum_{x(t-1)} W_{x(t) , x(t-1)}  P^*_{x(t-1)} \ln \left( W_{x(t) , x(t-1)} \right)
\nonumber \\
&& = \sum_{x} \sum_{y} W_{x , y} P^*_y  \ln \left( W_{x , y} \right) 
\label{p2pointsav}
\end{eqnarray}
which is independent of the time $t$ as a consequence of the stationarity of the dynamics.
So the average over trajectories of the information of Eq. \ref{infotrajchain}
reduces to
\begin{eqnarray}
< I[x(.)] >  
&& =  - \frac{ 1 }{T} \sum_{t=1}^T  < \ln \left( W_{x(t) , x(t-1)} \right) >   - \frac{ 1 }{T} < \ln \left(  P^*_{x(0)} \right) >
\nonumber \\
&& = -   \sum_{x} \sum_{y} W_{x , y} P^*_y  \ln \left( W_{x , y} \right)    - \frac{ 1 }{T} \sum_{x(0)}P^*_{x(0)}  \ln \left(  P^*_{x(0)}  \right) 
\label{infotrajchainav}
\end{eqnarray}
For $T \to +\infty$, the last contribution of order $1/T$ involving the initial condition disappears
and one recovers the standard expression for the Kolmogorov-Sinai entropy of Eq. \ref {hksdef}
 for discrete-time Markov chains with steady-state
\begin{eqnarray}
 h_{KS}=  \sum_{y} P^*_y  \left[ - \sum_{x} W_{x,y}   \ln \left(W_{x,y} \right) \right]
\label{hkschain}
\end{eqnarray}
So the Kolmogorov-Sinai entropy $h_{KS}$ can be explicitly computed in any model where the steady-state $P^*$ is known.


\subsection{ The scaled variance $V_{KS}$ of the information $I$ in terms of the temporal correlation of the flows }

Similarly for $T \to +\infty$, the contribution of the initial condition in the information of Eq. \ref{infotrajchain}
will disappear in the scaled variance $V_{KS}$ of Eq. \ref{psideri2} and one obtains
\begin{small}
  \begin{eqnarray}
&& V_{KS}= \oplim_{T \to +\infty} \left[   T \left( <I^2[x(.)]> - < I[x(.)]>^2 \right) \right] 
 =
 \oplim_{T \to +\infty} \left[   \frac{ 1 }{T }   \sum_{t=1}^T \left[ < \ln^2 \left( W_{x(t) , x(t-1)} \right) >
-   < \ln \left( W_{x(t) , x(t-1)} \right) >^2 \right] \right] 
 \nonumber \\ 
&& +  \oplim_{T \to +\infty} \bigg[  \frac{ 2 }{T } \sum_{t=1}^{T-1} \sum_{\tau=0}^{T-t-1} 
 \bigg( < \ln \left( W_{x(t+\tau+1) , x(t+\tau)} \right)  \ln \left( W_{x(t) , x(t-1)} \right) >
 -   < \ln \left( W_{x(t+\tau+1) , x(t+\tau)} \right) > <  \ln \left( W_{x(t) , x(t-1)} \right) > \bigg)
\bigg]
\nonumber 
\end{eqnarray}
\end{small}
The first line and the last term on the second line 
only involves the partial distribution of Eq. \ref{p2points}
for $N=2$ consecutive times,
while the first term on the second line involves the partial distribution of Eq. \ref{pNpoints}
for $N=4$ times, where the two first times and the two last times are consecutive,
while the intermediate time $\tau$ between the second time and the third time is arbitrary
\begin{eqnarray}
 p [x(t+\tau+1) ; x(t+\tau) ;x(t);x(t-1)]  = 
W_{x(t+\tau+1) , x(t+\tau)} \ \
\langle x(t+\tau) \vert W^{\tau} \vert x(t) \rangle \ \
W_{x(t) , x(t-1)} \ \ P^*(x(t-1))
\label{p4points}
\end{eqnarray}
Putting everything together, one obtains
 \begin{eqnarray}
V_{KS}
&& =
 \sum_{x,y}  W_{x,y} P^*_y  \ln^2 \left( W_{x ,y} \right) 
-   \left[ \sum_{x,y}   W_{x,y} P^*_y  \ln \left( W_{x ,y} \right)  \right]^2
 \nonumber \\  &&
 +    2   \sum_{x,y,x',y'}
  W_{x' , x} \ln \left( W_{x' , x} \right) 
  G_{x, y }
     W_{y,y'}   \ln \left( W_{y , y'} \right) P^*_{y'}
\label{varichainfinal}
\end{eqnarray}
where the notation $G_{x,y } $ represents
 the sum over the time $\tau$ of 
the difference between the finite-time propagator $\langle x \vert W^{\tau} \vert y \rangle  $ of Eq. \ref{chainpropa}
and its infinite-limit $P^*(x)$
\begin{eqnarray}
G_{x,y } \equiv \sum_{\tau=0}^{+\infty} \left[ \langle x \vert W^{\tau} \vert y \rangle -  P^*(x) \right]
= \sum_{\tau=0}^{+\infty} \sum_k e^{  - \tau \zeta_k } \langle x \vert \zeta^R_k \rangle \langle \zeta_k^L \vert y \rangle
 =  \sum_k \frac{ \langle x \vert \zeta^R_k \rangle \langle \zeta_k^L \vert y \rangle }{ 1- e^{  -  \zeta_k } }
\label{corre}
\end{eqnarray}
At the operator level, this Green function 
\begin{eqnarray}
G =  \sum_k \frac{  \vert \zeta^R_k \rangle \langle \zeta_k^L \vert }{ 1- e^{  -  \zeta_k } }
= \left( \sum_k  \vert \zeta^R_k \rangle \langle \zeta_k^L \vert   \right) 
\frac{\mathbb{1}}{  \mathbb{1} - W  }  \left( \sum_k  \vert \zeta^R_k \rangle \langle \zeta_k^L \vert   \right)
 = \left( \mathbb{1} - \vert r \rangle \langle l \vert  \right) 
\frac{\mathbb{1}}{  \mathbb{1} - W  }  \left( \mathbb{1} - \vert r \rangle \langle l \vert   \right)
\label{correop}
\end{eqnarray}
represents the inverse of the operator $(\mathbb{1} - W)$ within the subspace orthogonal to $\left(  \vert r \rangle \langle l \vert   \right) $.

So the explicit computation of the scaled variance $V_{KS}$ requires not only the knowledge of the steady-state $P^*$,
but also the knowledge of the Green function $G$.
This Green function is well-known in the perturbation theory of isolated eigenvalues (see Appendix \ref{app_per})
and appears more directly in the canonical analysis as explained below.
However, the more pedestrian derivation described above allows to see the link with the temporal correlations,
as explained previously for additive observables of Fokker-Planck dynamics in the context of Anderson localization \cite{ramola}.


\subsection{ Canonical analysis via the $\beta$-deformed Matrix ${\tilde W}^{[\beta]}_{x,y} $ }

Plugging the trajectories probabilities of Eq. \ref{pwtraj}
into the dynamical partition function of Eq. \ref{zbeta}
\begin{eqnarray}
Z_{\beta}(T)= \sum_{x(0 \leq t \leq T)}  \left( {\cal P}[x(0 \leq t \leq T)] \right)^{\beta}
= [P^*_{x(0)}]^{\beta}  \prod_{t=1}^T  \left[ W_{x(t) ,x(t-1)} \right]^{\beta} 
\opsimeq_{T \to +\infty}  e^{T \psi(\beta) }
\label{zbetachain}
\end{eqnarray}
yields that one needs to consider the $\beta$-deformed matrix
\begin{eqnarray}
{\tilde W}^{[\beta]}_{x,y} \equiv \left[ W_{x,y} \right]^{\beta} 
\label{wtildechain}
\end{eqnarray}
Then $e^{\psi(\beta)} $ corresponds to its highest eigenvalue that will dominate the deformed propagator for large $T$
\begin{eqnarray}
\langle x_T \vert \left({\tilde W}^{[\beta]} \right)^T \vert x_0 \rangle 
\opsimeq_{T \to + \infty} e^{ T \psi(\beta) } \  {\tilde r}^{[\beta]}_{x_T} \ {\tilde l}^{[\beta]}_{x_0}
\label{chainpropagator}
\end{eqnarray}
where ${\tilde r}^{[\beta]}_.$ and ${\tilde l}^{[\beta]}_.$ 
are the corresponding positive right and left eigenvectors of the Perron-Frobenius theorem
\begin{eqnarray}
e^{\psi(\beta)} {\tilde r}^{[\beta]}_x && = \sum_y {\tilde W}^{[\beta]}_{x,y}{\tilde r}^{[\beta]}_y 
= \sum_y \left[ W_{x,y} \right]^{\beta} {\tilde r}^{[\beta]}_y 
\nonumber \\
e^{\psi(\beta)} {\tilde l}_{\beta}(y) && = \sum_x {\tilde l}^{[\beta]}_x {\tilde W}^{[\beta]}_{x,y} 
=\sum_x {\tilde l}^{[\beta]}_x \left[ W_{x,y} \right]^{\beta}  
\label{tildechaineigen}
\end{eqnarray}
with the normalization
\begin{eqnarray}
\sum_x {\tilde l}^{[\beta]}_x  {\tilde r}^{[\beta]}_x =1
   \label{Wchainleftright}
\end{eqnarray}


\subsection{ Perturbation theory for the highest eigenvalue $e^{\psi(\beta=1+\epsilon)}$ at second order in $\epsilon$ }

The perturbation theory in $\beta=1+\epsilon$ of the deformed matrix of Eq. \ref{wtildechain}
\begin{eqnarray}
{\tilde W}^{[\beta=1+\epsilon]}_{x,y} = \left[ W_{x,y} \right]^{1+\epsilon} 
= W_{x,y} + \epsilon W^{(1)}_{x,y} + \epsilon^2 W^{(2)}_{x,y}+O(\epsilon^3)
\label{wtildechainper}
\end{eqnarray}
involves the first-order and the second-order perturbations
\begin{eqnarray}
 W^{(1)}_{x,y} && = W_{x,y} \ln \left( W _{x,y}\right)
 \nonumber \\
 W^{(2)}_{x,y} && = W_{x,y} \frac{ \ln^2 \left( W_{x,y}\right) }{2}
\label{wtildechainpere}
\end{eqnarray}

The perturbation theory for its highest eigenvalue
\begin{eqnarray}
e^{\psi(\beta=1+\epsilon)} = e^{\psi(1) +\epsilon \psi'(1)+ \frac{\epsilon^2}{2}  \psi''(1)+O(\epsilon^3)} 
= 1+ \epsilon \psi'(1)+ \frac{\epsilon^2}{2} \left[ \psi''(1) + [\psi'(1)]^2\right] +O(\epsilon^3)
\label{pereigenchain}
\end{eqnarray}
is recalled in Appendix \ref{app_per} and yields the following results at first-order and second-order respectively.

\subsubsection{ First-order perturbation to recover the Kolmogorov-Sinai entropy $h_{KS}=-\psi'(1)$ }

Using the unperturbed left and right eigenvectors of Eqs \ref{markovleft} and \ref{markovright},
one obtains that the first-order correction of Eq. \ref{energy1}
 for the eigenvalue of Eq. \ref{pereigenchain} reads
\begin{eqnarray}
\psi'(1) = \langle l \vert W^{(1)} \vert r \rangle
= \sum_{x,y} l_x W^{(1)}_{x,y} r_y = \sum_{x,y} W_{x,y} \ln \left( W_{x,y}\right) P^*_y
\label{perfirsth}
\end{eqnarray}
in agreement with the expression of Eq. \ref{hkschain}
for the Kolmogorov-Sinai entropy $h_{KS}=-\psi'(1)$.

\subsubsection{ Second-order perturbation to recover the scaled variance $V_{KS}=\psi''(1)$ }

The second-order correction of Eq. \ref{energy2}
 for the eigenvalue of Eq. \ref{pereigenchain} reads
in terms of the unperturbed left and right eigenvectors of Eqs \ref{markovleft} and \ref{markovright}
\begin{eqnarray}
  \frac{\psi''(1) +[\psi'(1)]^2}{2}  
&& =  \langle  l \vert  W^{(1)}  G W^{(1)}  \vert r \rangle
+\langle  l \vert  W^{(2)}  \vert r \rangle 
 =  \sum_{x,y,x',y'}   l_{x'} W^{(1)}_{x',x}  G_{x,y} W^{(1)}_{y,y'}  r_{y'}
+ \sum_{x,y} l_x W^{(2)}_{x,y} r_y
 \nonumber \\ 
 &&=  \sum_{x,y,x',y'}   W_{x',x} \ln \left( W_{x',x}\right)  G_{x,y} W_{y,y'} \ln \left( W_{y,y'}\right)  P^*_{y'}
+ \sum_{x,y}  W_{x,y} \frac{ \ln^2 \left( W_{x,y}\right) }{2} P^*_y
  \label{energy2chain}
\end{eqnarray}
where the Green function satisfies the matrix Eqs \ref{eqgreen} and \ref{eqgreenortho}
and thus coincides with Eq. \ref{correop}.
The equations \ref{eqgreen} and \ref{eqgreenortho} for the Green function read more explicitly in coordinates
\begin{eqnarray}
G_{x,y} - \sum_{x'} W_{x,x'}   G_{x',y} && =  \delta_{x,y} - P^*_x 
\nonumber \\
G_{x,y} -  \sum_{y'} G_{x,y'} W_{y',y}   && =   \delta_{x,y} - P^*_x 
\nonumber \\
\sum_{x}   G_{x,y} && =  0
\nonumber \\
\sum_{y}   G_{x,y} P^*_y && =  0
  \label{eqgreenchain}
\end{eqnarray}

The final result for $\psi''(1)$ 
\begin{eqnarray}
\psi''(1) 
 &&= 2 \sum_{x,y,x',y'}   W_{x',x} \ln \left( W_{x',x}\right)  G_{x,y} W_{y,y'} \ln \left( W_{y,y'}\right)  P^*_{y'}
\nonumber \\
&& + \sum_{x,y}  W_{x,y}  \ln^2 \left( W_{x,y}\right)  P^*_y
-    \left[ \sum_{x,y}   W_{x,y}   \ln \left( W_{x ,y} \right)  P^*_y\right]^2
  \label{psideri2chain}
\end{eqnarray}
coincides with Eq. \ref{varichainfinal} for the scaled variance $V_{KS}=\psi''(1)$
as it should (Eq. \ref{psideri2}).


\subsection { Conditioned process constructed via the generalization of Doob's h-transform}

The normalized probability to be at position $x$ at some interior time $0 \ll t \ll T$ for the dynamics
generated by the $\beta$-deformed matrix of Eq. \ref{wtildechain}
reads using the spectral asymptotic form of Eq. \ref{chainpropagator} for both time intervals $[0,t]$ and $[t,T]$
\begin{eqnarray}
{\tilde P}_x(t) 
&& 
 =  \frac{ \langle x_T \vert \left({\tilde W}^{[\beta]} \right)^{T-t} \vert x \rangle \langle x \vert \left({\tilde W}^{[\beta]} \right)^{t} \vert x_0 \rangle}
 { \displaystyle \sum_{x'}\langle x_T \vert \left({\tilde W}^{[\beta]} \right)^{T-t} \vert x' \rangle \langle x' \vert \left({\tilde W}^{[\beta]} \right)^{t} \vert x_0 \rangle} 
 \opsimeq_{ 0 \ll t \ll T }
 \frac{ e^{ (T-t) \psi(\beta) } {\tilde r}^{[\beta]}_{x_T} \  {\tilde l}^{[\beta]}_x 
 e^{ t \psi(\beta) } {\tilde r}^{[\beta]}_x \ {\tilde l}^{[\beta]}_{x_0} }
  {  \displaystyle \sum_{x'}  e^{ (T-t) \psi(\beta) } {\tilde r}^{[\beta]}_{x_T}\ {\tilde l}^{[\beta]}_{x'} e^{ t \psi(\beta) }
   {\tilde r}^{[\beta]}_{x'} \ {\tilde l}^{[\beta]}_{x_0}   } 
\nonumber \\
&&  \opsimeq_{ 0 \ll t \ll T } {\tilde l}^{[\beta]}_x \  {\tilde r}^{[\beta]}_x 
\label{conditionnedspectralint}
\end{eqnarray}
Since it is independent of the interior time $t$ as long as $0 \ll t \ll T$,
 it is useful to introduce the notation
\begin{eqnarray}
{\tilde {\tilde \rho}}^{[\beta]}_x \equiv  {\tilde l}^{[\beta]}_x  {\tilde r}^{[\beta]}_x 
 \label{rhokconditioned}
\end{eqnarray}
for the stationary density of the $\beta$-deformed dynamics in the interior time region $0 \ll t \ll T$,
and to construct the corresponding probability-preserving Markov matrix 
via the generalization of Doob's h-transform
\begin{eqnarray}
 {\tilde  {\tilde W}}^{[\beta]}_{x,y} = e^{ - \psi(\beta) } {\tilde l}^{[\beta]}_x  \  {\tilde W}^{[\beta]}_{x,y} \  \frac{1}{  {\tilde l}^{[\beta]}_y } 
  \label{Wdoobchain}
\end{eqnarray}
whose highest eigenvalue unity is associated to the trivial left eigenvector
\begin{eqnarray}
 {\tilde  {\tilde l}}^{[\beta]}_x = 1
   \label{Wdoobchainleft}
\end{eqnarray}
and to the right eigenvector $ {\tilde  {\tilde l}}^{[\beta]}_x ={\tilde {\tilde \rho}}^{[\beta]}_x $ of Eq. \ref{rhokconditioned}
that represents the normalized density conditioned to the information value
  $I_{\beta}=-\psi'(\beta)$ of the Legendre transform of Eq. \ref{legendrereci}.

So the explicit evaluation of the Doob generator of Eq. \ref{Wdoobchain}
requires the knowledge of the eigenvalue $ e^{  \psi(\beta) }$ and of the corresponding left eigenvector ${\tilde l}^{[\beta]}_. $
of Eq. \ref{tildechaineigen}.


\section{ Application to the discrete-time directed random trap model on the ring }

\label{sec_chaintrap}

In this section, the general analysis for discrete-time Markov chains described in the previous section
is applied to the directed trap model on the ring.

\subsection{ Model parametrization in terms of $L$ trapping times $\tau_y$ } 

The model is defined on a ring of $L$ sites with periodic boundary conditions $x+L \equiv x$,
and corresponds to the dynamics of Eq. \ref{markovchain}
where the
Markov Matrix
\begin{eqnarray}
W_{x,y} = \delta_{x,y+1}  \frac{ 1 }{\tau_y} + \delta_{x,y} \left(1- \frac{ 1 }{\tau_y}\right)  
\label{wchaintrapdirected}
\end{eqnarray}
involves the $L$ parameters $\tau_y >1$.
So when the particle is on site $y$ at time $t$,
it can either jump to the right neighbor $(y+1)$ with probability $\frac{ 1 }{\tau_y} \in ]0,1[$
or it remains on site $y$ with the complementary probability $\left(1- \frac{ 1 }{\tau_y}\right) \in ]0,1[$.
As a consequence, the escape-time $t$ from the site $y$ follows the geometric distribution for $t=1,2,...$
\begin{eqnarray}
p^{escape}_y(t) =   \frac{ 1 }{\tau_{y}}   \left(1- \frac{ 1 }{\tau_y}\right)^{t-1}
\label{escapechain}
\end{eqnarray}
whose averaged value is directly $\tau_y$
\begin{eqnarray}
\sum_{t=1}^{+\infty} t p^{escape}_y(t) =   \tau_{y}
\label{escapechainav}
\end{eqnarray}
So the $L$ parameters $\tau_y $ represent the characteristic times needed to escape from the $L$ sites $y=1,..,L$ of the ring. 


\subsection{ Minimal information $I_{min}$ and maximal information $I_{max}$ from extreme trajectories } 
 
Let us now consider some extreme trajectories.
The $L$ possible trajectories starting at $x(0) \in{1,2,..,L}$
and jumping forward at any time step $t=1,..,T$, making the large number $\frac{T}{L}$ of laps around the ring
have for probabilities
\begin{eqnarray}
  {\cal P}[x(t)=x(0)+t] \opsimeq_{T \to +\infty} \left( \prod_{y=1}^L \frac{1}{\tau_y} \right)^{\frac{T}{L}}
\label{minptrajex}
\end{eqnarray}
and correspond to the same intensive information that will be denoted by $I^{jump} $
\begin{eqnarray}
 I [x(t)=x(0)+t]= - \frac{1}{L} \sum_{y=1}^L \ln \left(\frac{ 1 }{\tau_{y }}\right) \equiv I^{jump}
\label{ijump}
\end{eqnarray}
On the contrary, the $L$ possible trajectories that remain on the same site $y$ of the ring for $0 \leq t \leq T$
have for probabilities
\begin{eqnarray}
  {\cal P}[x(t)= y ] = P^*_y \left(1- \frac{ 1 }{\tau_{y }}\right)^T
\label{maxptrajex}
\end{eqnarray}
and correspond to the different intensive informations
\begin{eqnarray}
  I[x(t)= y ] =  - \ln \left(1- \frac{ 1 }{\tau_{y }}\right) \equiv I^{loc}_y
\label{ilocy}
\end{eqnarray}
It is thus useful to introduce the site $y_{max}$ of the ring with the maximal trapping time 
and the site $y_{min}$ of the ring with the minimal trapping time 
\begin{eqnarray}
 \tau_{y_{max} } && = \max_{1 \leq y \leq L}\tau_y 
 \nonumber \\
  \tau_{y_{min} } && = \min_{1 \leq y \leq L}\tau_y 
\label{defymax}
\end{eqnarray}
To determine the minimal and the maximal informations, one should then distinguish three cases:

(i) If the probability $\left(1- \frac{ 1 }{\tau_y}\right)$ to remain on the site $y$ is always higher than 
the probability $\frac{ 1 }{\tau_y} $ to jump to the right neighbor $(y+1)$
\begin{eqnarray}
\frac{1}{   \tau_{y} } < \frac{1}{2} < 1- \frac{1}{   \tau_{y} } \ \ \ {\rm for } \ \ \ y=1,2,..,L
\label{case1}
\end{eqnarray}
then the maximal information will be given by Eq. \ref{ijump},
while the minimal information will be given by Eq. \ref{ilocy} for $y=y_{max}$
\begin{eqnarray}
 I^{max} && = I^{jump} = \frac{1}{L} \sum_{y=1}^L \ln(\tau_y)
 \nonumber \\
 I^{min} && =  I^{loc}_{y_{max}}= - \ln  \left(1- \frac{ 1 }{\tau_{y_{max} }}\right)
\label{icase1}
\end{eqnarray}

(ii) If the probability $\left(1- \frac{ 1 }{\tau_y}\right)$ to remain on the site $y$ is always smaller than 
the probability $\frac{ 1 }{\tau_y} $ to jump to the right neighbor $(y+1)$
\begin{eqnarray}
1- \frac{1}{   \tau_{y} } < \frac{1}{2} <  \frac{1}{   \tau_{y} } \ \ \ {\rm for } \ \ \ y=1,2,..,L
\label{case2}
\end{eqnarray}
then the maximal information will be given by Eq. \ref{ilocy} for $y=y_{min}$,
while the minimal information will be given by Eq. \ref{ijump}
\begin{eqnarray}
 I^{max} && = I^{loc}_{y_{min}}= - \ln  \left(1- \frac{ 1 }{\tau_{y_{min} }}\right)
 \nonumber \\
 I^{min} && = I^{jump} = \frac{1}{L} \sum_{y=1}^L \ln(\tau_y)
\label{icase2}
\end{eqnarray}

(iii) In the remaining cases
\begin{eqnarray}
\frac{1}{   \tau_{y_{max}} } < \frac{1}{2} <  \frac{1}{   \tau_{y_{min}} } 
\label{case3}
\end{eqnarray}
the maximal information will be given by Eq. \ref{ilocy} for $y=y_{min}$,
while the minimal information will be given by Eq. \ref{ilocy} for $y=y_{max}$
\begin{eqnarray}
 I^{max} && = I^{loc}_{y_{min}}= - \ln  \left(1- \frac{ 1 }{\tau_{y_{min} }}\right)
 \nonumber \\
 I^{min} && =  I^{loc}_{y_{max}} = - \ln  \left(1- \frac{ 1 }{\tau_{y_{max} }}\right)
\label{icase3}
\end{eqnarray}


\subsection{ Explicit results for the Kolmogorov-Sinai entropy $h_{KS}$ } 

The normalized steady state of Eq. \ref{markovchainst}
\begin{eqnarray}
P^*_x =  \sum_y W_{x,y} P^*_y= 
 \frac{ 1 }{\tau_{x-1}} P^*_{x-1}+  \left(1- \frac{ 1 }{\tau_x}\right) P^*_x
\label{markovchainstring}
\end{eqnarray}
is simply given by the weight of the trapping time $\tau_y$ within the sum of all the trapping times of the ring
\begin{eqnarray}
 P^*_y  = \frac{ \tau_y } { \displaystyle \sum_{x=1}^{L} \tau_{x}}
\label{chaintrapstsol}
\end{eqnarray}
As a consequence, the Kolmogorov-Sinai entropy of Eq. \ref{hkschain} reads
for a given disordered ring parametrized by the $L$ trapping times $\tau_{y=1,2,..,L}$
\begin{eqnarray}
 h_{KS}[\tau_{y=1,2,..,L} ]  && =  \sum_{y=1}^L P^*_y  \left[ -  W_{y+1,y}   \ln \left(W_{y+1,y} \right)-  W_{y,y}   \ln \left(W_{y,y} \right) \right]
  =  \frac{\displaystyle \sum_{y=1}^L \tau_y  \left[ - \frac{1}{\tau_y}   \ln \left( \frac{1}{\tau_y} \right) 
  -   \left(1-  \frac{1}{\tau_y} \right)     \ln \left(1- \frac{1}{\tau_y}  \right) \right] } 
{ \displaystyle \sum_{x=1}^{L} \tau_{x}}
\nonumber \\
&&  =  \frac{\displaystyle \sum_{y=1}^L  \left[     \ln \left( \tau_y \right) 
  -   \left(\tau_y -  1 \right)     \ln \left(1- \frac{1}{\tau_y}  \right) \right] } 
{ \displaystyle \sum_{x=1}^{L} \tau_{x}}
\label{hkschaintrap}
\end{eqnarray}

Let us now analyze its behavior for large $L$ when the probability distribution 
$q(\tau)$ of the trapping times $\tau \in ]1,+\infty[$
is the power-law of Eq. \ref{qtaupower}
depending on the parameter $\mu>0$ : 

(i) in the region $\mu>1$ where the averaged value $\overline{\tau} $ of the trapping time is finite (Eq. \ref{mom1}),
both the numerator and the denominator of Eq. \ref{hkschaintrap} will follow the law of large numbers
and the Kolmogorov-Sinai entropy will converge towards the finite asymptotic value 
\begin{eqnarray}
 h_{KS}^{(L=\infty)}  =   \frac{\displaystyle \int_1^{+\infty} d \tau  q(\tau)  \left[     \ln \left( \tau \right) 
  -   \left(\tau -  1 \right)     \ln \left(1- \frac{1}{\tau}  \right) \right] } 
{ \displaystyle \int_1^{+\infty} d \tau \tau q(\tau)}
\ \ \ \ \ {\rm for}  \ \ \ \mu >1
\label{hkschaintraptypfinite}
\end{eqnarray}

(ii) in the region $0<\mu<1$ where the averaged value $\overline{\tau} $ of the trapping time is infinite (Eq. \ref{mom1}),
the numerator of Eq. \ref{hkschaintrap} will still follow the law of large numbers,
while the denominator is a L\'evy sum that remains distributed as recalled in  Appendix \ref{app_levy}.
As a consequence, the Kolmogorov-Sinai entropy will not remain finite as in Eq. \ref{hkschaintraptypfinite},
but will vanish with the scaling $L^{1-\frac{1}{\mu} } $
\begin{eqnarray}
 h_{KS}^{(L)}  && \opsimeq_{L \to +\infty} L^{1-\frac{1}{\mu} }  \ \frac{1}{\theta}  \int_1^{+\infty} d \tau  q(\tau)  \left[     \ln \left( \tau \right) 
  -   \left(\tau -  1 \right)     \ln \left(1- \frac{1}{\tau}  \right) \right] 
  \ \ \ \ \ {\rm for}  \ \ \ 0<\mu<1
  \label{hkschaintraptyplevy1}
\end{eqnarray}
and will remain distributed over the disordered rings of length $L$ since the rescaled variable $\theta$ of Eq. \ref{thetalevy1}
is distributed with the L\'evy law ${\cal L}_{\mu}(\theta)$ of index $\mu \in ]0,1[$ of Eq. \ref{levy1laplace}.


\subsection{ Canonical analysis via the $\beta$-deformed Markov Matrix }

For the Markov matrix of Eq. \ref{wchaintrapdirected},
the $\beta$-deformed matrix of Eq. \ref{wtildechain} reads
\begin{eqnarray}
{\tilde W}^{[\beta]}_{x,y}  \equiv \left[ W_{x,y} \right]^{\beta} 
=\delta_{x,y+1}  \left(\frac{ 1 }{\tau_y } \right)^{\beta }+ \delta_{x,y} \left(1- \frac{ 1 }{\tau_y}\right)^{\beta }
\label{wtildechaintrap}
\end{eqnarray}
and the corresponding eigenvalues Eqs \ref{tildechaineigen} become
\begin{eqnarray}
e^{\psi(\beta)} {\tilde r}^{[\beta]}_x &&   
= \sum_y \left[ \delta_{x,y+1}  \left(\frac{ 1 }{\tau_y } \right)^{\beta } 
+ \delta_{x,y} \left(1- \frac{ 1 }{\tau_y}\right)^{\beta}   \right] {\tilde r}^{[\beta]}_y
=   \frac{ 1 }{\tau^{\beta}_{x-1}}  {\tilde r}^{[\beta]}_{x-1} 
+ \left(1- \frac{ 1 }{\tau_x}\right)^{\beta}   {\tilde r}^{[\beta]}_x
\nonumber \\
e^{\psi(\beta)} {\tilde l}^{[\beta]}_y &&  
=\sum_x {\tilde l}^{[\beta]}_x \left[ \delta_{x,y+1}  \left(\frac{ 1 }{\tau_y } \right)^{\beta } 
+ \delta_{x,y} \left(1- \frac{ 1 }{\tau_y}\right)^{\beta}   \right] 
=  \frac{ 1 }{\tau^{\beta}_y}  {\tilde l}^{[\beta]}_{y+1} 
+ \left(1- \frac{ 1 }{\tau_y}\right)^{\beta}  {\tilde l}^{[\beta]}_y  
\label{tildechaineigentrap}
\end{eqnarray}
The solutions of these recursions read
\begin{eqnarray}
 {\tilde r}^{[\beta]}_x &&   
=   \frac{ 1 }{\tau^{\beta}_{x-1}\left[ e^{\psi(\beta)}-  \left(1- \frac{ 1 }{\tau_x}\right)^{\beta} \right]}  {\tilde r}^{[\beta]}_{x-1}
= {\tilde r}^{[\beta]}_0 \prod_{y=1}^x
 \frac{ 1 }{\tau^{\beta}_{y-1}\left[ e^{\psi(\beta)}-  \left(1- \frac{ 1 }{\tau_y}\right)^{\beta} \right]} 
\nonumber \\
 {\tilde l}^{[\beta]}_y  && =
  \tau^{\beta}_{y-1}\left[ e^{\psi(\beta)} -\left(1- \frac{ 1 }{\tau_{y-1}}\right)^{\beta} \right] {\tilde l}^{[\beta]}_{y-1} 
 = {\tilde l}^{[\beta]}_0  \prod_{x=1}^y  \tau^{\beta}_{x-1}\left[ e^{\psi(\beta)} -\left(1- \frac{ 1 }{\tau_{x-1}}\right)^{\beta} \right] 
\label{tildechaineigentrapsol}
\end{eqnarray}
The periodic boundary conditions ${\tilde r}^{[\beta]}_L = {\tilde r}^{[\beta]}_0$ and $ {\tilde l}^{[\beta]}_L = {\tilde l}^{[\beta]}_0  $
yield the equation for the eigenvalue $e^{\psi(\beta)}$
\begin{eqnarray}
1 =  \prod_{x=1}^L \left( \tau^{\beta}_{x}\left[ e^{\psi(\beta)} -\left(1- \frac{ 1 }{\tau_{x}}\right)^{\beta} \right] \right)
=  \prod_{x=1}^L  \left[ e^{\psi(\beta)} \tau^{\beta}_{x}-\left(\tau_{x}-1\right)^{\beta} \right] 
\label{psiringchaintrap}
\end{eqnarray}
while the positivity of the components of the Perron-Froebenius eigenvectors of Eqs \ref{tildechaineigentrap}
imply
\begin{eqnarray}
e^{\psi(\beta)} \geq \left(1- \frac{ 1 }{\tau_{x}}\right)^{\beta} \ \ \ {\rm for } \ \ \ x=1,2,..,L
\label{psiringchaintrappos}
\end{eqnarray}


\subsection { Corresponding conditioned process constructed via the generalization of Doob's h-transform}

Using the left eigenvector of Eq. \ref{tildechaineigentrap},
one obtains that probability-preserving Markov matrix 
obtained via the generalization of Doob's h-transform of Eq. \ref{Wdoobchain}
is of the same form of the initial Markov matrix of Eq. \ref{wchaintrapdirected}
\begin{eqnarray}
 {\tilde  {\tilde W}}^{[\beta]}_{x,y} = e^{ - \psi(\beta) } {\tilde l}^{[\beta]}_x  \  {\tilde W}^{[\beta]}_{x,y} \  \frac{1}{  {\tilde l}^{[\beta]}_y } 
 =  \delta_{x,y+1}  \frac{ 1 }{{\tilde  {\tilde \tau}}_{\beta}(y)} + \delta_{x,y} \left(1- \frac{ 1 }{{\tilde  {\tilde \tau}}_{\beta}(y)}\right)  
  \label{Wdoobchaintrap}
\end{eqnarray}
where the modified trapping time ${\tilde  {\tilde \tau}}^{[\beta]}_y $ at position $y$
depends on the initial trapping time $\tau_y $ at position $y$, on $\beta$ and the eigenvalue $\psi(\beta)$
\begin{eqnarray}
\frac{ 1 }{{\tilde  {\tilde \tau}}^{[\beta]}_y}  = 1- e^{ - \psi(\beta) }   \left(1- \frac{ 1 }{\tau_y}\right)^{\beta }
  \label{Wdoobchaineff}
\end{eqnarray}
The corresponding conditioned density of Eq. \ref{rhokconditioned}
is given by the analog of the steady state of Eq. \ref{chaintrapstsol}
with the modified trapping times of Eq. \ref{Wdoobchaineff}
\begin{eqnarray}
{\tilde {\tilde \rho}}^{[\beta]}_y  = \frac{ {\tilde  {\tilde \tau}}^{[\beta]}_y } { \displaystyle \sum_{x=1}^{L}{\tilde  {\tilde \tau}}^{[\beta]}_x}
\label{chaintrapdoobrho}
\end{eqnarray}
Let us now describe special values of $\beta$.


\subsection{ Special value $\beta=0$ } 

For $\beta=0$, Eq. \ref{psiringchaintrap} leads to the simple value independent of the trapping times
\begin{eqnarray}
    e^{\psi(\beta=0)} =2
\label{psiringchaintrapzero}
\end{eqnarray}
as it should to reproduce the total number
\begin{eqnarray}
    Z_T(\beta=0)=2^T
\label{zchaintrapzero}
\end{eqnarray}
of possible trajectories of $T$ steps
of Eq. \ref{zbetazero} for the present model, where there are two possibilities at each step (Eq. \ref{wchaintrapdirected}).
The modified trapping times of Eq. \ref{Wdoobchaineff} become all equal to $2$
\begin{eqnarray}
\frac{ 1 }{{\tilde  {\tilde \tau}}^{[\beta=0]}_y}  = \frac{1}{2}
  \label{Wdoobchaineffzero}
\end{eqnarray}
as it should to have equal probabilities $\left(\frac{1}{2},\frac{1}{2} \right) $ for the two possibilities to jump or to remain on site.
Accordingly, the corresponding conditioned density of Eq. \ref{chaintrapdoobrho} becomes uniform
\begin{eqnarray}
{\tilde {\tilde \rho}}^{[\beta=0]}_y  = \frac{ 1}{L}
\label{chaintrapdoobrhozero}
\end{eqnarray}


\subsection{ Limit $\beta \to +\infty$ and the minimal intensive information $I_{min} $ } 

In the limit $\beta \to +\infty$, one expects that $\psi(\beta)$ is negative with the linear behavior of Eq. \ref{zbetainfty}
\begin{eqnarray}
 \psi(\beta)  \opsimeq_{\beta \to + \infty} - \beta I_{min}  
\label{zbetainftyp}
\end{eqnarray}
Then the condition of Eq. \ref{psiringchaintrappos} yields
\begin{eqnarray}
 I_{min} \leq -  \ln \left(1- \frac{ 1 }{\tau_{x}}\right) \ \ \ {\rm for } \ \ \ x=1,2,..,L
\label{psiringchaintrapposmin}
\end{eqnarray}
while
Eq. \ref{psiringchaintrap} becomes
\begin{eqnarray}
0 \opsimeq_{\beta \to + \infty}  \sum_{x=1}^L \ln \left( \tau^{\beta}_{x}
\left[ e^{- \beta I_{min}} - e^{\beta \ln \left(1- \frac{ 1 }{\tau_{x}}\right)} \right] \right)
\opsimeq_{\beta \to + \infty}  
\beta \sum_{x=1}^L \ln \left( \tau_{x} \right)
+ \sum_{x=1}^L \ln \left[ e^{- \beta I_{min}} - e^{\beta \ln \left(1- \frac{ 1 }{\tau_{x}}\right)} \right] 
\label{psiringchaintrappi}
\end{eqnarray}
so that one needs to distinguish whether the inequality of Eq. \ref {psiringchaintrapposmin} is strict or not.

\subsubsection{ Case where the inequality of Eq. \ref {psiringchaintrapposmin} remains strict} 

If the inequality of Eq. \ref{psiringchaintrapposmin} remains strict
\begin{eqnarray}
 I_{min} < -  \ln \left(1- \frac{ 1 }{\tau_{y_{max}}}\right) 
\label{psiringchaintrapposminstrict}
\end{eqnarray}
then Eq. \ref{psiringchaintrappi} leads to the solution
\begin{eqnarray}
I_{min} = \frac{1}{L} \sum_{x=1}^L \ln \left( \tau_{x} \right) \equiv I^{jump}
\label{psiringchaintrappisolstrict}
\end{eqnarray}
corresponding to the value $I^{jump}$ of Eq. \ref{ijump}.
This solution is valid only if $I^{jump} $ satisfies the strict bound of Eq. \ref{psiringchaintrapposminstrict}
\begin{eqnarray}
I^{jump} \equiv \frac{1}{L} \sum_{x=1}^L \ln \left( \tau_{x} \right)  < -  \ln \left(1- \frac{ 1 }{\tau_{y_{max}}}\right) 
\label{psiringchaintrapposminstrictcondition}
\end{eqnarray}

In the Doob generator of the conditioned process,
the modified trapping times of Eq. \ref{Wdoobchaineff} become all equal to unity
\begin{eqnarray}
{\tilde  {\tilde \tau}}_{\beta}(y) 
\opsimeq_{\beta \to + \infty} \frac{ 1 }{1- e^{  \beta \left[ I^{jump}  + \ln \left(1- \frac{ 1 }{\tau_y}\right) \right] } }
\opsimeq_{\beta \to + \infty} 1
  \label{Wdoobchaineffstrict}
\end{eqnarray}
as it should to have probability one to jump and probability zero to remain on site.
Accordingly, 
the conditioned density of Eq. \ref{chaintrapdoobrho} becomes uniform
\begin{eqnarray}
{\tilde {\tilde \rho}}_{\beta} (y)  \opsimeq_{\beta \to + \infty} \frac{ 1}{L}
\label{chaintrapdoobrhozeroinftim}
\end{eqnarray}


\subsubsection{ Case where the inequality of Eq. \ref {psiringchaintrapposmin} cannot remain strict } 

If $I^{jump}$ does not satisfy the inequality of Eq. \ref{psiringchaintrapposminstrictcondition},
then the solution of Eq. \ref{psiringchaintrappi}
is instead
\begin{eqnarray}
 I^{min} =  - \ln  \left(1- \frac{ 1 }{\tau_{y_{max}}}\right) \equiv I^{loc}_{y_{max}}
\label{iminexrec}
\end{eqnarray}
corresponding to the value $ I^{loc}_{y_{max}}$ discussed in Eqs \ref{ilocy} and \ref{defymax}.

In the Doob generator of the conditioned process,
the modified trapping time of Eq. \ref{Wdoobchaineff} 
\begin{eqnarray}
{\tilde  {\tilde \tau}}^{[\beta]}_y  \opsimeq_{\beta \to + \infty} 
\frac{ 1 }{ 1- e^{ \beta \left[  I^{loc}_{y_{max}}  +   \ln \left(1- \frac{ 1 }{\tau_y}\right) \right] } }
 =  \frac{ 1 }{1- e^{ \beta \left[  - \ln  \left(1- \frac{ 1 }{\tau_{y_{max}}}\right) +  \ln \left(1- \frac{ 1 }{\tau_y}\right) \right] } }
 && \opsimeq_{\beta \to + \infty}  1 \ \ \ {\rm if } \ \ y \ne y_{max}
 \nonumber \\
 && \opsimeq_{\beta \to + \infty}  + \infty \ \ \ {\rm if } \ \ y = y_{max}
  \label{Wdoobchaineffinftymax}
\end{eqnarray}
remains finite for $y \ne y_{max}$ but diverges for $y=y_{max}$,
so that the corresponding conditioned density of Eq. \ref{chaintrapdoobrho}
is fully localized on the site $y_{max} $
\begin{eqnarray}
{\tilde {\tilde \rho}}^{[\beta]}_y  \opsimeq_{\beta \to + \infty}  \delta_{y,y_{max}}
\label{chaintrapdoobrhoinftypmax}
\end{eqnarray}


\subsection{ Series expansion in $\beta=1+\epsilon$ up to order $\epsilon^2$ } 

For $\beta=1+\epsilon$, the expansion of the logarithm of Eq. \ref{psiringchaintrap} reads up to second order in $\epsilon$ 
\begin{eqnarray}
&& 0  =  \sum_{x=1}^L \ln \left[ e^{\psi(1+\epsilon)} \tau^{1+\epsilon}_{x} -\left(\tau_{x}-1\right)^{1+\epsilon} \right] 
=  \sum_{x=1}^L \ln \left[ \tau_x e^{\epsilon \left[ \psi'(1)+\ln (\tau_x) \right] + \frac{\epsilon^2}{2} \psi''(1) }  
- (\tau_{x}-1 ) e^{\epsilon \ln (\tau_{x}-1 ) } \right] 
\nonumber \\
&& 
=  \sum_{x=1}^L \ln \left[ 
1 
+ \epsilon \left[ \tau_x  \left[ \psi'(1)+\ln (\tau_x) \right]  -  (\tau_{x}-1 )\ln (\tau_{x}-1 )\right]
+ \frac{\epsilon^2}{2} \left[ \tau_x \left[ \psi''(1)+\left[ \psi'(1)+\ln (\tau_x) \right]^2 \right] - (\tau_{x}-1 )\ln^2 (\tau_{x}-1 ) \right]
  \right]   
  \nonumber \\
&& 
=   \epsilon \sum_{x=1}^L \left[ \tau_x   \psi'(1)+\tau_x \ln (\tau_x)   -  (\tau_{x}-1 )\ln (\tau_{x}-1 )\right]
 + \frac{\epsilon^2}{2} 
\sum_{x=1}^L  
\left(  \tau_x \psi''(1) -  \tau_x  (\tau_{x}-1 ) \left[ \psi'(1)+\ln (\tau_x) - \ln (\tau_{x}-1 )\right]^2
\right)
\label{psiringchaintrapeps}
\end{eqnarray}
So the vanishing at order $\epsilon$ yields the first derivative
\begin{eqnarray}
\psi'(1) = -  \frac{\displaystyle \sum_{x=1}^L  \left[ \tau_x   \ln \left(\tau_x \right) -   \left(\tau_x-1\right)     \ln \left( \tau_x-1 \right) \right] } 
{ \displaystyle \sum_{y=1}^{L} \tau_{y}}
\label{firstderi}
\end{eqnarray}
in agreement with $h_{KS}=-\psi'(1)$ of Eq. \ref{hkschaintrap}.
The vanishing at order $\epsilon^2$ yields the second derivative $\psi''(1)$ 
\begin{eqnarray}
 \psi''(1) =
\frac{  \displaystyle  \sum_{x=1}^L   \tau_x ( \tau_x -1) \left[ \psi'(1) + \ln (\tau_x)  - \ln (\tau_{x}-1 ) \right]^2 }
{ \displaystyle \sum_{y=1}^{L} \tau_{y}}
\label{secondderi}
\end{eqnarray}
The scaled variance $V_{KS}$ of Eq. \ref{psideri2}
thus reads
for a given disordered ring parametrized by the $L$ trapping times $\tau_{y=1,2,..,L}$
\begin{eqnarray}
 && V_{KS}[\tau_{y=1,2,..,L} ]  = 
\frac{  \displaystyle  \sum_{x=1}^L   \tau_x ( \tau_x -1) \left[ - h_{KS}[\tau_.]   - \ln \left( 1- \frac{1}{\tau_x}\right) \right]^2 }
{ \displaystyle \sum_{y=1}^{L} \tau_{y}}
\nonumber \\
&& =\frac{  \displaystyle   h_{KS}^2[\tau_.] \sum_{x=1}^L    ( \tau^2_x - \tau_x)    
+ 2 h_{KS}[\tau_.]  \sum_{x=1}^L    ( \tau^2_x - \tau_x)     \ln \left( 1- \frac{1}{\tau_x}\right) 
+ \sum_{x=1}^L   ( \tau^2_x - \tau_x)    \ln^2 \left( 1- \frac{1}{\tau_x}\right) }
{ \displaystyle \sum_{y=1}^{L} \tau_{y}}
\label{varkstrapchain}
\end{eqnarray}
where $h_{KS}[\tau_{y=1,2,..,L} ] $ was given in Eq. \ref{hkschaintrap}.

Let us now analyze its behavior for large $L$ when the probability distribution 
$q(\tau)$ of the trapping times $\tau \in ]1,+\infty[$
is the power-law of Eq. \ref{qtaupower}
depending on the parameter $\mu>0$ : 

(i) in the region $\mu>2$ where the second moment $\overline{\tau^2} $ of the trapping time is finite (Eq. \ref{mom2}),
both the numerator and the denominator of Eq. \ref{varkstrapchain} will follow the law of large numbers
while the Kolmogorov-Sinai entropy converges towards the finite asymptotic value of Eq. \ref{hkschaintraptypfinite}.
As a consequence, the scaled variance of Eq. \ref{varkstrapchain} will then 
converges towards the finite asymptotic value
\begin{eqnarray}
&& V_{KS}^{(\infty)}   =   \frac{\displaystyle 
\int_1^{+\infty} d \tau  q(\tau)    \tau ( \tau -1) \left[ - h_{KS}^{(\infty)}  - \ln \left( 1- \frac{1}{\tau}\right) \right]^2 } 
{ \displaystyle \int_1^{+\infty} d \tau \tau q(\tau)}
\ \ \ \ \ {\rm for}  \ \ \ \mu >2
\label{vkschaintraptypfinite} \\
&& = \frac{\displaystyle 
\left[  h_{KS}^{(\infty)}   \right]^2\int_1^{+\infty} d \tau  q(\tau)     ( \tau^2 -\tau) 
+2 h_{KS}^{(\infty)}   \int_1^{+\infty} d \tau  q(\tau)   ( \tau^2 -\tau)    \ln \left( 1- \frac{1}{\tau}\right) 
+ \int_1^{+\infty} d \tau  q(\tau)    ( \tau^2 -\tau)   \ln^2 \left( 1- \frac{1}{\tau}\right)  } 
{ \displaystyle \int_1^{+\infty} d \tau \tau q(\tau)}
\nonumber
\end{eqnarray}

(ii) in the region $1<\mu<2$ where the second moment $\overline{\tau^2} $ of the trapping time is infinite (Eq. \ref{mom2})
while the first moment $\overline{\tau} $ remains finite  (Eq. \ref{mom1}),
the only anomalous scaling will come from the sum of the square of the trapping times 
discussed around Eq. \ref{upsilon}.
As a consequence, the scaled variance $V_{KS}$ will not remain finite as in Eq. \ref{vkschaintraptypfinite},
but will diverge with the scaling $L^{\frac{2}{\mu} -1 } $ of exponent $\left(\frac{2}{\mu} -1 \right) \in ]0,1[ $ 
\begin{eqnarray}
 && V_{KS}^{(L)}  \opsimeq_{L \to + \infty}  L^{\frac{2}{\mu} -1 } \ \vartheta \ 
 \frac{ \left[  h_{KS}^{(\infty)}   \right]^2   } 
{ \displaystyle \int_1^{+\infty} d \tau \tau q(\tau)}
\ \ \ \ \ {\rm for}  \ \ \ 1<\mu <2
\label{varkstrapchainlevy12}
\end{eqnarray}
and it will remain distributed over the disordered rings of length $L$ since the rescaled variable $\vartheta$ of Eq. \ref{thetalevy1sq}
is distributed with the L\'evy law ${\cal L}_{\frac{\mu}{2}}(\vartheta) $.

(iii) in the region $0<\mu<1$ where both the first moment $\overline{\tau} $
and the second moment $\overline{\tau^2} $ are infinite (Eqs \ref {mom1} \ref{mom2})
while the Kolmogorov-Sinai entropy does not converge anymore towards the finite asymptotic value of Eq. \ref{hkschaintraptypfinite}, one needs to return to the finite-size expression of Eq. \ref{hkschaintrap}
for the Kolmogorov-Sinai entropy
and to re-analyze the leading behavior of Eq. \ref{varkstrapchain}
in terms of the sum $\Sigma_L$ of Eq. \ref{sigmaq} and $\Upsilon_L$ of Eq. \ref{upsilon}
\begin{eqnarray}
 V_{KS}[\tau_{y=1,2,..,L} ]  \opsimeq_{L \to + \infty} 
 \frac{  \displaystyle     \Upsilon_L     }  
{ \displaystyle \Sigma_L^3} L^2  \left( \int_1^{+\infty} d \tau  q(\tau)  \left[     \ln \left( \tau \right) 
  -   \left(\tau -  1 \right)     \ln \left(1- \frac{1}{\tau}  \right) \right] \right)^2
  \oppropto_{L \to + \infty } L^{2-\frac{1}{\mu}  }
  \ \ \ \ {\rm for}  \ \ 0<\mu <1
\label{varkstrapchainlevy01}
\end{eqnarray}
so the scaling in $ L^{2-\frac{1}{\mu}  } $ is different from Eq. \ref{varkstrapchainlevy12},
while the limit distribution would require a more refined analysis of the ratio $ \frac{     \Upsilon_L     }  
{  \Sigma_L^3}  $ involving the two correlated sums of Eq. \ref{sigmaq} and  Eq. \ref{upsilon}.


\subsection{Direct analysis of self-averaging observables in the thermodynamic limit of an infinite ring $L \to +\infty$ } 

As discussed above, the Kolmogorov-Sinai entropy $h_{KS}=-\psi'(1)$ 
is self-averaging in the thermodynamic limit $L \to +\infty$ only for $\mu>1$ (Eq. \ref{hkschaintraptypfinite}),
while the scaled variance $V_{KS}=\psi''(1)$ is self-averaging in the thermodynamic limit $L \to +\infty$ only for $\mu>2$
(Eq. \ref{vkschaintraptypfinite}). Further transitions are expected for the higher cumulants.

However if the trapping time distribution $q(\tau)$ has all its moments finite (in contrast to the power-law form of Eq. \ref{qtaupower}
discussed up to now), then the scaled cumulant generating function $\psi(\beta)$ 
and its derivative will be self-averaging in the thermodynamic limit $L \to +\infty$.
If one rewrites Eq. \ref{psiringchaintrap} via its logarithm and divide by the size $L$ of the ring
\begin{eqnarray}
0 = \frac{1}{L}  \sum_{x=1}^L  \ln \left[ e^{\psi(\beta)} \tau^{\beta}_{x}-\left(\tau_{x}-1\right)^{\beta} \right] 
\label{psiringchaintraplog}
\end{eqnarray}
one obtains that the self-averaging value $\psi_{L=\infty}(\beta)$ in the thermodynamic limit $L \to +\infty$ 
is determined by the equation
\begin{eqnarray}
0 = \int_1^{+\infty} d\tau q(\tau)   \ln \left[ e^{\psi_{\infty}(\beta)} \tau^{\beta}-\left(\tau-1\right)^{\beta} \right] 
\label{psithermolim}
\end{eqnarray}
However, whenever there are non-self-averaging effects, one should return to the finite-size Eq. \ref{psiringchaintraplog}
to analyze them, as described above for the two first derivatives $\psi'(1) = -h_{KS}$ and $\psi''(1)=V_{KS}$.


\section{ Markov Jump Processes in continuous time with steady-state }

\label{sec_jump}

In this section, we focus on the continuous-time dynamics in discrete space defined by the Master Equation
\begin{eqnarray}
\frac{\partial P_x(t)}{\partial t} =    \sum_{y }   w_{x,y}  P_y(t)
\label{mastereq}
\end{eqnarray}
where the off-diagonal $x \ne y$ matrix elements are positive $w_{x,y} \geq 0 $  and represent the transitions rates from $y$ to $x $,
while the diagonal elements are negative $w_{x,x} \leq 0 $ and are fixed by the conservation of probability to be
\begin{eqnarray}
w_{x,x} && =  - \sum_{y \ne x} w_{y,x} 
\label{wdiag}
\end{eqnarray}

\subsection{ Steady-State and finite-time propagator }

We will assume that
the normalized steady-state $P^*(x)$ of Eq. \ref{mastereq}
\begin{eqnarray}
0 =    \sum_{y }   w_{x,y}  P^*_y = \sum_{y \ne x  }\left[    w_{x,y}  P^*_y -  w_{y,x}  P^*_x \right]
\label{mastereqst}
\end{eqnarray}
exists.
Eqs \ref{wdiag} and \ref{mastereqst} mean that zero is the highest eigenvalue of the Markov Matrix $w_{.,.}$,
with the positive left eigenvector 
\begin{eqnarray}
 l_x=1
\label{markovleftj}
\end{eqnarray}
and the positive right eigenvector $r_x$ given by the steady state
\begin{eqnarray}
 r_x=P^*_x
\label{markovrightj}
\end{eqnarray}
The whole spectral decomposition of the matrix $w$
\begin{eqnarray}
 w && = -  \sum_k \zeta_k \vert \zeta^R_k \rangle \langle\zeta_k^L  \vert 
\label{jumpspectral}
\end{eqnarray}
involving the other eigenvalues $(-\zeta_k)<0$ labelled by $k$,
with their right eigenvectors $\vert \zeta^R_k \rangle $ and their left eigenvectors $ \langle \zeta_k^L  \vert  $
satisfying the closure relation
\begin{eqnarray}
 \mathbb{1} && = \vert r \rangle \langle l \vert + \sum_k  \vert \zeta^R_k \rangle \langle\zeta_k^L  \vert
\label{fermeturej}
\end{eqnarray}
is useful to describe the relaxation of the finite-time propagator towards the steady state
\begin{eqnarray}
\langle x \vert e^{wt} \vert x_0 \rangle = 
\langle x \vert r \rangle \langle l \vert x_0 \rangle+ \sum_k e^{  -  t \zeta_k }
\langle x \vert \zeta^R_k \rangle \langle\zeta_k^L  \vert x_0 \rangle
= P^*(x) + \sum_k e^{  - t \zeta_k } \langle x \vert \zeta^R_k \rangle \langle\zeta_k^L  \vert x_0 \rangle 
\label{jumppropa}
\end{eqnarray}


\subsection{ The intensive information $I$ as a time-additive observable of the trajectory $x(0 \leq t \leq T) $ }

A dynamical trajectory $x(t)$ on the time interval $0 \leq t \leq T$
corresponds to a certain number $M \geq 0 $ of jumps $m=1,..,M$ occurring at times $0<t_1<...<t_M<T$
between the successive configurations $(x_0 \to x_1 \to x_2.. \to x_M)$ that are visited between these jumps.
The probability density of this trajectory
\begin{eqnarray}
x(0 \leq t \leq T) = \left( x_0; t_1 ; x_1; t_2 ;... ; x_{M-1} ; t_{M} ; x_M \right)
\label{traject}
\end{eqnarray}
when the initial condition $x_0 $ is drawn with the steady-state distribution $P^*$ of Eq. \ref{mastereqst}
reads in terms of the transitions rates 
\begin{eqnarray}
&& {\cal P}[x(0 \leq t \leq T)=\left( x_0; t_1 ; x_1; t_2 ;... ; x_{M-1} ; t_{M} ; x_M \right) ]
\nonumber \\
&& = 
 e^{ (T-t_M) w_{x_M,x_M} } 
 w_{x_M , x_{M-1} } 
 e^{ (t_M-t_{M-1} ) w_{x_{M-1},x_{M-1}} } 
 ...
  .... w_{x_{2} , x_{1} } 
 e^{ (t_2-t_1) w_{x_1,x_1} } 
 w_{x_1 ,x_{0} }
 e^{ t_1 w_{x_0,x_0} }  
 \nonumber \\
&& = P^*(x_0) e^{ (T-t_M) w_{x_M,x_M} } \prod_{m=1}^M \left[ w_{x_m , x_{m-1} } e^{ (t_m-t_{m-1} ) w_{x_{m-1},x_{m-1}} } \right]
\label{ptraject}
\end{eqnarray}

The normalization over all possibles trajectories on $[0,T]$
involves the sum over the number $M$ of jumps, the sum 
over the $M$ configurations $(x_0,x_1,...,x_M$)
where $x_m$ has to be different from $x_{m-1}$, 
and the integration over the jump times with the measure $dt_1... dt_M$ and the constraint $0<t_1<...<t_M<T $
\begin{eqnarray}
 1  = && \sum_{M=0}^{+\infty}  \int_0^T dt_M \int_0^{t_M} dt_{M-1} ... \int_0^{t_2} dt_{1} 
 \sum_{x_M \ne x_{M-1}}\sum_{x_{M-1} \ne x_{M-2}}  
...
\sum_{x_2 \ne x_1}  \sum_{x_1 \ne x_0} \sum_{x_0}
 \nonumber \\ && 
 {\cal P}[x(0 \leq t \leq T)=\left( x_0; t_1 ; x_1; t_2 ;... ; x_{M-1} ; t_{M} ; x_M \right) ]
\label{norma}
\end{eqnarray}
The trajectory probability density of Eq. \ref{ptraject}
 can be rewritten more compactly without the explicit enumeration of all the jumps as
\begin{eqnarray}
{\cal P}[x(0 \leq t \leq T)]   
=  P^*_{x(0)} e^{ \displaystyle  \left[  \sum_{t: x(t^+) \ne x(t^-) } \ln ( w_{x(t^+) , x(t^-) } ) +  \int_0^T dt  w_{x(t) , x(t) }  \right]  }
\label{pwtrajjump}
\end{eqnarray}
The corresponding intensive information of Eq. \ref{infotraj}
\begin{eqnarray}
I[x(0 \leq t \leq T)] 
=   - \frac{1}{T} \sum_{t: x(t^+) \ne x(t^-) } \ln ( w_{x(t^+) , x(t^-) } )  - \frac{1}{T}  \int_0^T dt  w_{x(t) , x(t) }
  - \frac{ 1 }{T}\ln \left(  P^*_{x(0)}  \right)
\label{infotrajjump}
\end{eqnarray}
is a time-additive observable.
As a consequence, its averaged value with the trajectories probabilities of Eq. \ref{pwtrajjump}
reads
\begin{eqnarray}
< I[x(.)] > && \equiv \sum_{x(.)} {\cal P}[x(.)]  I[x(.)]  
 =   - \frac{1}{T} \sum_{t: x(t^+) \ne x(t^-) } < \ln ( w_{x(t^+) , x(t^-) } ) >  - \frac{1}{T}  \int_0^T dt  < w_{x(t) , x(t) } >
  - \frac{ 1 }{T} < \ln \left(  P^*_{x(0)}  \right) >
  \nonumber \\
&&=  -  \sum_{y  }\sum_{ x \ne y  }  w_{x,y}  P^*_y  \ln ( w_{x,y} ) - \sum_y   P^*_y     w_{y,y}   
   - \frac{ 1 }{T} \sum_{x(0)}P^*_{x(0)}  \ln \left(  P^*_{x(0)} \right)
\label{infotrajchainempilimit}
\end{eqnarray}
For $T \to +\infty$, the last contribution of order $1/T$ involving the initial condition disappears
and one obtains the Kolmogorov-Sinai entropy of Eq. \ref {hksdef}
that can be rewritten 
in terms of the off-diagonal elements $ w_{x,y}$ alone using Eq. \ref{wdiag}
\begin{eqnarray}
h_{KS} =    \oplim_{T \to +\infty} \left( < I[x(.)] > \right) 
&& =  -  \sum_{y  }\sum_{ x \ne y  }  w_{x,y}  P^*_y  \ln ( w_{x,y} ) - \sum_y   P^*_y  \left( - \sum_{x \ne y } w_{x,y} \right)
 \nonumber \\ &&
 = \sum_y   P^*_y     \sum_{x \ne y} w_{x,y} \left[ 1      -     \ln ( w_{x,y} ) \right]
\label{hksjump}
\end{eqnarray}
As in Eq. \ref{hkschain}, the Kolmogorov-Sinai entropy can be thus explicitly computed in any model where the steady-state $P^*$ is known.


\subsection{ The scaled variance $V_{KS}$ of the information $I$ in terms of the temporal correlations }

Similarly for $T \to +\infty$, the contribution of the initial condition in the information of Eq. \ref{infotrajjump}
will disappear in the scaled variance of Eq. \ref{psideri2} and one obtains
the contributions corresponding to the various connected temporal correlations 
involving either two off-diagonal matrix elements, one off-diagonal and one diagonal matrix elements, or two diagonal matrix elements
\begin{small}
  \begin{eqnarray}
&& V_{KS} = \oplim_{T \to +\infty} \left[   T \left( <I^2[x(.)]> - < I[x(.)]>^2 \right) \right] 
 =
 \oplim_{T \to +\infty} \left[  
    \frac{1}{T} \sum_{t: x(t^+) \ne x(t^-) }  <\ln^2 ( w_{x(t^+) , x(t^-) } )  >
  \right] 
  \nonumber \\ 
  && 
  \! \!  \!  \! \!  \!  \! \!  \! 
  +
 \oplim_{T \to +\infty} \left[ 
 \frac{2}{T}  \! \!  \!  \! \!  \!  \! \!  
 \sum_{  \substack{t: x(t^+) \ne x(t^-) \\ \tau>0: x((t+\tau)^+) \ne x((t+\tau)^-)}  } 
 \! \!  \!  \! \!  \!  \! \!  \!  \! \!
 \left[< \ln ( w_{x((t+\tau)^+) , x((t+\tau)^-) } ) \ln ( w_{x(t^+) , x(t^-) } ) >
 - < \ln ( w_{x((t+\tau)^+) , x((t+\tau)^-) } )> <  \ln ( w_{x(t^+) , x(t^-) } ) >
\right]  \right] 
\nonumber \\ 
&& +  \oplim_{T \to +\infty} \bigg[  
\frac{2}{T} \sum_{t: x(t^+) \ne x(t^-) } \int_0^{T-t} d\tau 
\left[ < w_{x(t+\tau) , x(t+\tau) } \ln ( w_{x(t^+) , x(t^-) } ) >
- < w_{x(t+\tau) , x(t+\tau) } > < \ln ( w_{x(t^+) , x(t^-) } ) >
\right]
\bigg]
\nonumber \\ 
&&
 \! \!  \!  \! \!  \!  \! \!  \!  +  \oplim_{T \to +\infty} \bigg[  
\frac{2}{T}   \int_0^T dt  
 \! \!  \!  \! \!  \!  \! \!  \! 
 \sum_{0<\tau<T-t: x((t+\tau)^+) \ne x((t+\tau)^-) }
  \! \!  \!  \! \!  \!  \! \!  \! 
 \left[ < \ln (  w_{x((t+\tau)^+) , x((t+\tau)^-) } )w_{x(t) , x(t) } >
 - < \ln (  w_{x((t+\tau)^+) , x((t+\tau)^-) } ) > < w_{x(t) , x(t) } >\right]
\bigg]
\nonumber \\ 
&& +  \oplim_{T \to +\infty} \bigg[  
 \frac{2}{T}  \int_0^T dt \int_0^{T-t} d\tau \left[ <w_{x(t+\tau) , x(t+\tau) } w_{x(t) , x(t) } >
 - <w_{x(t+\tau) , x(t+\tau) } > < w_{x(t) , x(t) } >
 \right]
\bigg]
\label{varijump}
\end{eqnarray}
\end{small}

As in Eq. \ref{corre}, it is useful to introduce the notation $G_{x,y } $
for the integral over the time $\tau$ of 
the difference between the finite-time propagator $\langle x \vert e^{\tau w} \vert y \rangle  $ of Eq. \ref{jumppropa}
and its infinite-limit $P^*(x)$
\begin{eqnarray}
G_{x,y } \equiv \int_{0}^{+\infty} d \tau \left[ \langle x \vert e^{\tau w} \vert y \rangle -  P^*(x) \right]
= \int_{0}^{+\infty} d \tau \sum_k e^{  - \tau \zeta_k } \langle x \vert \zeta^R_k \rangle \langle \zeta_k^L \vert y \rangle
 =  \sum_k \frac{ \langle x \vert \zeta^R_k \rangle \langle \zeta_k^L \vert y \rangle }{  \zeta_k  }
\label{correjump}
\end{eqnarray}
At the operator level, this Green function 
\begin{eqnarray}
G =  \sum_k \frac{  \vert \zeta^R_k \rangle \langle \zeta_k^L \vert }{  \zeta_k  }
= \left( \sum_k  \vert \zeta^R_k \rangle \langle \zeta_k^L \vert   \right) 
\frac{\mathbb{1}}{  -w  }  \left( \sum_k  \vert \zeta^R_k \rangle \langle \zeta_k^L \vert   \right)
 = \left( \mathbb{1} - \vert r \rangle \langle l \vert  \right) 
\frac{\mathbb{1}}{  (-w)  }  \left( \mathbb{1} - \vert r \rangle \langle l \vert   \right)
\label{correopjump}
\end{eqnarray}
represents the inverse of the operator $(-w)$ within the subspace orthogonal to $\left(  \vert r \rangle \langle l \vert   \right) $.

Putting everything together, the scaled variance of Eq. \ref{varijump} reads
  \begin{eqnarray}
 V_{KS} 
&& = \sum_y \sum_{x \ne y}  w_{x,y} P^*_y  \ln^2 \left( w_{x ,y} \right) 
  \nonumber \\ 
  && 
 +    2   \sum_{x,y} \sum_{x' \ne x} \sum_{y' \ne y}
  w_{x' , x} \ln \left( w_{x' , x} \right) 
  G_{x, y }
     w_{y,y'}   \ln \left( w_{y , y'} \right) P^*_{y'} 
\nonumber \\ 
&& 
 +   2    \sum_{x,y}  \sum_{y' \ne y}
  w_{x , x}  
  G_{x, y }
     w_{y,y'}   \ln \left( w_{y , y'} \right) P^*_{y'} 
\nonumber \\ 
&& 
+   2    \sum_{x,y} \sum_{x' \ne x} 
  w_{x' , x} \ln \left( w_{x' , x} \right) 
  G_{x, y }
     w_{y,y}    P^*_{y} 
\nonumber \\ 
&&   +    2   \sum_{x,y} 
  w_{x , x}  
  G_{x, y }
     w_{y,y}  P^*_{y} 
\label{varijumpfinal}
\end{eqnarray}
As in Eq. \ref{varichainfinal},
the explicit computation of the scaled variance $V_{KS}$ requires not only the knowledge of the steady-state $P^*$,
but also the knowledge of the Green function $G$, that appears more directly in the canonical analysis below.


\subsection{ Canonical analysis via the $\beta$-deformed Matrix $ {\tilde w}^{[\beta]}_{x,y}$ }

Plugging the trajectory probability density of Eq. \ref{pwtrajjump}
into the dynamical partition function of Eq. \ref{zbeta}
\begin{eqnarray}
\left( {\cal P}[x(0 \leq t \leq T)]   \right)^{\beta}
= \left[  P^*_{x(0)}  \right]^{\beta} e^{ \displaystyle  \left[  \sum_{t: x(t^+) \ne x(t^-) } \ln ( \left[ w_{x(t^+) , x(t^-) }\right]^{\beta} ) +  \int_0^T dt  \beta w_{x(t) , x(t) }  \right]  }
\label{pwtrajjumpbeta}
\end{eqnarray}
yields that one needs to consider the $\beta$-deformed matrix
with the off-diagonal elements  
\begin{eqnarray}
  {\tilde w}^{[\beta]}_{x,y} && = \left[ w_{x,y}\right]^{\beta} \ \ \  {\rm for } \ \ \ x \ne y
\label{jumptilt}
\end{eqnarray}
and with the diagonal elements using Eq. \ref{wdiag}
\begin{eqnarray}
  {\tilde w}^{[\beta]}_{y,y} && = \beta  w_{y,y} = - \beta \sum_{x \ne y} w_{x,y}
\label{jumptiltdiag}
\end{eqnarray}

Then $\psi(\beta) $ corresponds to its highest eigenvalue that will dominate the deformed propagator for large $T$
\begin{eqnarray}
\langle x_T \vert e^{T {\tilde w}^{[\beta]} } \vert x_0 \rangle 
\opsimeq_{T \to + \infty} e^{ T \psi(\beta) } {\tilde r}^{[\beta]}_{x_T} {\tilde l}^{[\beta]}_{x_0}
\label{jumppropagator}
\end{eqnarray}
with the corresponding positive right and left eigenvectors of the Perron-Frobenius theorem
\begin{eqnarray}
\psi(\beta) {\tilde r}^{[\beta]}_x && = \sum_y {\tilde w}^{[\beta]}_{x,y} {\tilde r}^{[\beta]}_y 
=  \sum_{y \ne x} 
\left(  \left[ w_{x,y} \right]^{\beta} {\tilde r}^{[\beta]}_y  
-  \beta  w_{y,x}  {\tilde r}^{[\beta]}_x \right)
\nonumber \\
\psi(\beta) {\tilde l}^{[\beta]}_y && = \sum_x {\tilde l}^{[\beta]}_x {\tilde w}^{[\beta]}_{x,y}
= \sum_{x \ne y} \left(  {\tilde l}^{[\beta]}_x  \left[ w_{x,y} \right]^{\beta}
- {\tilde l}^{[\beta]}_x  \beta  w_{x,y} \right)
\label{tildejumpeigen}
\end{eqnarray}
with the normalization
\begin{eqnarray}
\sum_x {\tilde l}^{[\beta]}_x  {\tilde r}^{[\beta]}_x =1
   \label{Wdoobjumpleftright}
\end{eqnarray}


\subsection{ Perturbation theory for the eigenvalue $\psi(\beta=1+\epsilon)$ at second order in $\epsilon$ }

The perturbation theory in $\beta=1+\epsilon$ of the deformed matrix of Eqs \ref{jumptilt} and \ref{jumptiltdiag}
\begin{eqnarray}
{\tilde w}^{[\beta=1+\epsilon]}_{x,y}  = w_{x,y} + \epsilon w^{(1)}_{x,y} + \epsilon^2 w^{(2)}_{x,y}+O(\epsilon^3)
\label{wtildejumpper}
\end{eqnarray}
involves the first-order matrix elements
\begin{eqnarray}
 w^{(1)}_{x,y} && = w_{x,y} \ln \left( w_{x,y} \right) \ \ \  {\rm for } \ \ \ x \ne y
 \nonumber \\
 w^{(1)}_{y,y} &&  = w_{y,y}  = -  \sum_{x \ne y} w(x,y)
\label{wtildejump1}
\end{eqnarray}
and the second-order matrix elements
\begin{eqnarray}
 w^{(2)}_{x,y} && = w_{x,y} \frac{\ln^2 \left( w_{x,y}\right)}{2}  \ \ \  {\rm for } \ \ \ x \ne y
 \nonumber \\
 w^{(2)}_{y,y} &&  = 0
\label{wtildejump2}
\end{eqnarray}

The perturbation theory for the highest eigenvalue
\begin{eqnarray}
\psi(\beta=1+\epsilon) = \psi(1) +\epsilon \psi'(1)+ \frac{\epsilon^2}{2}  \psi''(1) +O(\epsilon^3)
\label{pereigenjump}
\end{eqnarray}
is recalled in Appendix \ref{app_per} with the following results at first order ans second order respectively.


\subsubsection{ First-order perturbation theory to recover the Kolmogorov-Sinai entropy $h_{KS}=-\psi'(1)$ }

Using the unperturbed left and right eigenvectors of Eqs \ref{markovleftj} and \ref{markovrightj},
one obtains that the first-order correction of Eq. \ref{energy1}
 for the eigenvalue of Eq. \ref{pereigenjump} reads
\begin{eqnarray}
\psi'(1) && =  \langle l \vert w^{(1)} \vert r \rangle
= \sum_{x,y} l_x w^{(1)}_{x,y} r_y
= \sum_{y} \left[ \sum_{x\ne y} l_x w^{(1)}_{x,y} r_y + l_y w^{(1)}_{y,y} r_y \right]
\nonumber \\
&& = \sum_{y}  \left[ \sum_{x \ne y } w_{x,y} \ln \left( w _{x,y}\right) + w_{y,y}\right] P^*_y
=
\sum_{y}  \left[ \sum_{x \ne y } w_{x,y} \ln \left( w _{x,y}\right) -  \sum_{x \ne y} w_{x,y}\right] P^*_y
\label{perfirsthj}
\end{eqnarray}
in agreement with the expression of Eq. \ref{hksjump}
for the Kolmogorov-Sinai entropy $h_{KS}=-\psi'(1)$.

\subsubsection{ Second-order perturbation theory to recover the scaled variance $V_{KS}=\psi''(1)$ }

The second-order correction of Eq. \ref{energy2}
 for the eigenvalue of Eq. \ref{pereigenjump} reads
in terms of the unperturbed left and right eigenvectors of Eqs \ref{markovleftj} and \ref{markovrightj}
\begin{eqnarray}
  \frac{\psi''(1)}{2}  
&& = \langle  l \vert  w^{(2)}  \vert r \rangle 
+ \langle  l \vert  w^{(1)}  G w^{(1)}  \vert r \rangle
 =  \sum_{x,y} l_x w^{(2)}_{x,y} r_y
 + \sum_{x,x',y',y}   l_{x'} w^{(1)}_{x',x}  G_{x,y} w^{(1)}_{y,y'}  r_{y'}
 \nonumber \\ 
&& =  \sum_{y}  \sum_{x \ne y }  w_{x,y} \frac{\ln^2 \left( w_{x,y} \right)}{2} P^*_y
+ \sum_{x,x',y',y}  w^{(1)}_{x',x}  G_{x,y} w^{(1)}_{y,y'}  P^*_{y'}
    \label{energy2jump}
\end{eqnarray}
where the Green function satisfies the matrix Eqs \ref{eqgreen} and \ref{eqgreenortho}
and thus coincides with Eq. \ref{correopjump}.
The Equations \ref{eqgreen} and \ref{eqgreenortho}
for the Green function read more explicitly in coordinates
\begin{eqnarray}
 - \sum_{x'} w_{x,x'}   G_{x',y} && =  \delta_{x,y} - P^*_x 
\nonumber \\
-  \sum_{y'} G_{x,y'} w_{y',y}  && =   \delta_{x,y} - P^*_x 
\nonumber \\
\sum_{x}   G_{x,y} && =  0
\nonumber \\
\sum_{y}   G_{x,y} P^*_y && =  0
  \label{eqgreenjump}
\end{eqnarray}
Using
the diagonal and off-diagonal elements of the first-order perturbation matrix $w^{(1)}$ of Eq. \ref{wtildejump1},
one obtains that the final result for $\psi''(1)$ of Eq. \ref{energy2jump}
\begin{eqnarray}
 \psi''(1)  
&& =   \sum_{y}  \sum_{x \ne y }  w_{x,y} \ln^2 \left( w_{x,y} \right) P^*_y
  \nonumber \\ 
  &&  +2  \sum_{x,y} \sum_{x' \ne x} \sum_{y' \ne y}  w_{x',x} \ln \left( w_{x',x} \right)  G_{x,y} w_{y,y'} \ln \left( w_{y,y'} \right)  P^*_{y'}
  \nonumber \\ 
  &&   +2  \sum_{x,y}  \sum_{y' \ne y}  w_{x,x}  G_{x,y} w_{y,y'} \ln \left( w_{y,y'} \right) P^*_{y'}
  \nonumber \\ 
  &&    +2 \sum_{x,y} \sum_{x' \ne x} \  w_{x',x} \ln \left( w_{x',x} \right)  G_{x,y} w_{y,y}  P^*_{y}
  \nonumber \\ 
  &&     +2  \sum_{x,y}   w_{x,x}  G_{x,y} w_{y,y}  P^*_{y}
    \label{energy2deriful}
\end{eqnarray}
coincides with Eq. \ref{varijumpfinal} for the scaled variance $V_{KS}=\psi''(1)$ as it should.


\subsection { Corresponding conditioned process constructed via the generalization of Doob's h-transform}

The analysis analogous to Eq. \ref{conditionnedspectralint}
yields that 
\begin{eqnarray}
{\tilde {\tilde \rho}}^{[\beta]}_x \equiv  {\tilde l}^{[\beta]}_x \  {\tilde r}^{[\beta]}_x
 \label{rhokconditionedj}
\end{eqnarray}
represents the stationary density of the $\beta$-deformed dynamics in the interior time region $0 \ll t \ll T$
and can be interpreted as the normalized density conditioned to the information value
  $I=-\psi'(\beta)$ of the Legendre transform of Eq. \ref{legendrereci}.

The corresponding probability-preserving Markov jump process
whose highest eigenvalue is zero with the corresponding trivial left eigenvector $ {\tilde  {\tilde l}}_{\beta}(x) = 1$
and the corresponding right eigenvector given by Eq. \ref{rhokconditionedj}
is generated by the following matrix 
 corresponding to the generalization of Doob's h-transform
\begin{eqnarray}
 {\tilde  {\tilde w}}^{[\beta]}_{x,y} = {\tilde l}^{[\beta]}_x \  {\tilde w}^{[\beta]}_{x,y} \  \frac{1}{  {\tilde l}^{[\beta]}_y} - \psi(\beta) \delta_{x,y}
  \label{wdoobjump}
\end{eqnarray}
So the explicit evaluation of this Doob generator
requires the knowledge of the eigenvalue $   \psi(\beta) $ and of the corresponding left eigenvector ${\tilde l}^{[\beta]}_. $
of Eq. \ref{tildejumpeigen}.


\section{ Application to the continuous-time directed random trap model  }

\label{sec_jumptrap}

In this section, the general analysis for continuous-time Markov jump processes described in the previous section
is applied to the directed trap model on the ring.

\subsection{ Model parametrization in terms of $L$ trapping times $\tau_y$ } 

The model is defined on a ring of $L$ sites with periodic boundary conditions $x+L \equiv x$,
and corresponds to the dynamics of Eq. \ref{mastereq}
where the Markov Matrix
\begin{eqnarray}
w_{x,y} =   \frac{ \delta_{x,y+1} - \delta_{x,y}}{\tau_y}
\label{wjumptrap}
\end{eqnarray}
involves the $L$ parameters $\tau_y >0$.
So when the particle is on site $y$ at time $t$,
it can either jump to the right neighbor $(y+1)$ with rate $\frac{ 1 }{\tau_y}$ per unit time,
or it remains on site $y$.
As a consequence, the escape-time $t \in [0,+\infty[$ from the site $y$ follows the exponential distribution
\begin{eqnarray}
p^{escape}_y(t) =   \frac{ 1 }{\tau_{y}}  e^{ - \frac{ t }{\tau_y} }
\label{escapejump}
\end{eqnarray}
whose averaged value is directly $\tau_y$
\begin{eqnarray}
\int_0^{+\infty}dt  t p^{escape}_y(t) =   \tau_{y}
\label{escapejumpav}
\end{eqnarray}
So the $L$ parameters $\tau_y $ represent again the characteristic times needed to escape from the $L$ sites $y=1,..,L$ of the ring. 


\subsection{ Minimal information $I_{min}$ and maximal information $I_{max}$ from extreme trajectories }

 The $L$ possible trajectories that remain on the same site $y$ of the ring for $0 \leq t \leq T$
have for probabilities
\begin{eqnarray}
  {\cal P}[x(t)= y ] =  P^*_y e^{ \displaystyle - \frac{ T }{\tau_{y }}}
\label{maxptrajexw}
\end{eqnarray}
and correspond to the different intensive informations
\begin{eqnarray}
  I[x(t)= y ] =   \frac{ 1 }{\tau_{y }} \equiv I^{loc}_y
\label{ilocyw}
\end{eqnarray}
In terms of the positions $y_{max}$ and $y_{min} $ of the ring with the maximal and the minimal trapping time 
(Eq \ref{defymax}),
 the maximal information and the minimal information are thus given by
\begin{eqnarray}
 I^{max} && = I^{loc}_{y_{min}}=  \frac{ 1 }{\tau_{y_{min} }}
  \nonumber \\
 I^{min} && =  I^{loc}_{y_{max}} =  \frac{ 1 }{\tau_{y_{max} }}
\label{icase3w}
\end{eqnarray}


\subsection{ Explicit results for the Kolmogorov-Sinai entropy $h_{KS}$ } 

The normalized steady state of Eq. \ref{mastereqst}
\begin{eqnarray}
0  =  \frac{ 1 }{\tau_{x-1}} P^*_{x-1}- \frac{ 1 }{\tau_x} P^*_x
\label{mastereqsttrap}
\end{eqnarray}
is simply given by
\begin{eqnarray}
 P^*_y  = \frac{ \tau_y } { \displaystyle \sum_{x=1}^{L} \tau_{x}}
\label{jumptrapstsol}
\end{eqnarray}
As a consequence, the Kolmogorov-Sinai entropy of Eq. \ref{hksjump} reads
for a given disordered ring parametrized by the $L$ trapping times $\tau_{y=1,2,..,L}$
\begin{eqnarray}
h_{KS} [\tau_{y=1,2,..,L} ] &&  =  \sum_y   P^*_y      w_{y+1,y} \left[ 1      -     \ln ( w_{y+1,y} ) \right]
=\frac{\displaystyle \sum_{y=1}^{L}    \left[ 1      +     \ln ( \tau_y ) \right] } 
{ \displaystyle \sum_{x=1}^{L} \tau_{x}}
= \frac{\displaystyle L + \sum_{y=1}^{L}        \ln ( \tau_y )  } 
{ \displaystyle \sum_{x=1}^{L} \tau_{x}}
\label{hksjumptrap}
\end{eqnarray}

Let us now analyze its behavior for large $L$ when the probability distribution 
$q(\tau)$ of the trapping times $\tau \in ]1,+\infty[$
is the power-law of Eq. \ref{qtaupower}
depending on the parameter $\mu>0$ : 

(i) in the region $\mu>1$ where the averaged value $\overline{\tau} $ of the trapping time is finite (Eq. \ref{mom1}),
both the numerator and the denominator of Eq. \ref{hksjumptrap} will follow the law of large numbers
and the Kolmogorov-Sinai entropy will converge towards the finite asymptotic value 
\begin{eqnarray}
 h_{KS}^{(L=\infty) } =   \frac{\displaystyle 1+ \int_1^{+\infty} d \tau  q(\tau)       \ln \left( \tau \right)  } 
{ \displaystyle \int_1^{+\infty} d \tau \tau q(\tau)}
\ \ \ \ \ {\rm for}  \ \ \ \mu >1
\label{hksjumptraptypfinite}
\end{eqnarray}

(ii) in the region $0<\mu<1$ where the averaged value $\overline{\tau} $ of the trapping time is infinite (Eq. \ref{mom1}),
the numerator of Eq. \ref{hksjumptrap} will still follow the law of large numbers,
while the denominator is a L\'evy sum that remains distributed as recalled in  Appendix \ref{app_levy}.
As a consequence, the Kolmogorov-Sinai entropy will not remain finite as in Eq. \ref{hksjumptraptypfinite},
but will vanish with the scaling $L^{1-\frac{1}{\mu} } $
\begin{eqnarray}
 h_{KS}^{(L) } && \opsimeq_{L \to +\infty} L^{1-\frac{1}{\mu} }  \ \frac{1}{\theta} 
\left[  1+ \int_1^{+\infty} d \tau  q(\tau)       \ln \left( \tau \right) \right]
  \ \ \ \ \ {\rm for}  \ \ \ 0<\mu<1
  \label{hksjumptraptyplevy1}
\end{eqnarray}
and will remain distributed over the disordered rings of length $L$ since the rescaled variable $\theta$ of Eq. \ref{thetalevy1}
is distributed with the L\'evy law ${\cal L}_{\mu}(\theta)$ of index $\mu \in ]0,1[$ of Eq. \ref{levy1laplace}.


\subsection{ Canonical analysis via the $\beta$-deformed Markov Matrix }

For the Markov matrix of Eq. \ref{wjumptrap},
the $\beta$-deformed matrix of Eqs \ref{jumptilt}
and \ref{jumptiltdiag}
 reads
\begin{eqnarray}
  {\tilde w}^{[\beta]}_{x,y} && = \delta_{x,y+1}  \left(\frac{ 1 }{\tau_y } \right)^{\beta }-  \delta_{x,y} \frac{ \beta }{\tau_y }
\label{jumptilttrap}
\end{eqnarray}
and the corresponding eigenvalues Eqs \ref{tildejumpeigen} become
\begin{eqnarray}
\psi(\beta)  {\tilde r}^{[\beta]}_x && 
=   \frac{ {\tilde r}^{[\beta]}_{x-1} }{\tau^{\beta} _{x-1}}  
-   \frac{  \beta }{\tau_x}   {\tilde r}^{[\beta]}_x
\nonumber \\
\psi(\beta) {\tilde l}^{[\beta]}_y && 
=    \frac{{\tilde l}^{[\beta]}_{y+1}   }{\tau^{\beta}_y} 
-    \frac{ \beta }{\tau_y} {\tilde l}^{[\beta]}_y
\label{tildejumpeigentrap}
\end{eqnarray}
The solutions of these recursions read
\begin{eqnarray}
 {\tilde r}^{[\beta]}_x  
&& =   \frac{ {\tilde r}^{[\beta]}_{x-1} }{\tau^{\beta}_{x-1}\left[ \psi(\beta) + \frac{ \beta  }{\tau_x}  \right] }  
=  \frac{ {\tilde r}^{[\beta]}_0 }{ \displaystyle \prod_{y=1}^x
\left( \tau^{\beta}_{y-1}\left[ \psi(\beta) + \frac{ \beta  }{\tau_y}  \right] \right) }   
\nonumber \\
{\tilde l}^{[\beta]}_y  && =\tau^{\beta}_{y-1} \left[ \psi(\beta) + \frac{\beta}{\tau_{y-1} } \right]  {\tilde l}^{[\beta]}_{y-1}  
=  {\tilde l}^{[\beta]}_0  \prod_{x=1}^y \left( \tau^{\beta}_{x-1} \left[ \psi(\beta) + \frac{\beta}{\tau_{x-1} } \right] \right)
\label{tildejumpeigentrapsol}
\end{eqnarray}
The periodic boundary conditions ${\tilde r}^{[\beta]}_L = {\tilde r}^{[\beta]}_0$ and $ {\tilde l}^{[\beta]}_L = {\tilde l}^{[\beta]}_0  $
yield the equation for the eigenvalue $\psi(\beta)$
\begin{eqnarray}
1 && =  \prod_{x=1}^L \left( \tau^{\beta}_{x} \left[ \psi(\beta) + \frac{\beta}{\tau_{x} } \right] \right)
=  \prod_{x=1}^L  \left[ \psi(\beta) \tau^{\beta}_{x}+ \beta \tau^{\beta-1}_{x} \right] 
\label{psijumpeigentrapsol}
\end{eqnarray}
while the positivity of the components of the Perron-Froebenius eigenvectors of Eqs \ref{tildejumpeigentrapsol}
imply
\begin{eqnarray}
\psi(\beta) \geq - \frac{ \beta }{\tau_{x}} \ \ \ {\rm for } \ \ \ x=1,2,..,L
\label{psiringjumptrappos}
\end{eqnarray}


\subsection { Corresponding conditioned process constructed via the generalization of Doob's h-transform}

Using the left eigenvector of Eq. \ref{tildejumpeigentrapsol},
one obtains that probability-preserving Markov matrix 
obtained via the generalization of Doob's h-transform of Eq. \ref{wdoobjump}
is of the same form of the initial Markov matrix of Eq. \ref{wjumptrap}
\begin{eqnarray}
{\tilde  {\tilde w}}^{[\beta]}_{x,y} =  {\tilde l}^{[\beta]}_x   {\tilde w}^{[\beta]}_{x,y}  \frac{1}{  {\tilde l}^{[\beta]}_y } - \psi(\beta) \delta_{x,y}
= \frac{ \delta_{x,y+1} - \delta_{x,y}}{ {\tilde  {\tilde \tau}}_{\beta}(y) }
\label{wjumptrapdoob}
\end{eqnarray}
with the modified trapping times
\begin{eqnarray}
\frac{ 1 }{{\tilde  {\tilde \tau}}^{[\beta]}_y}  = \psi(\beta) + \frac{ \beta }{\tau_y}
  \label{Wdoobjumpeff}
\end{eqnarray}
The corresponding conditioned density of Eq. \ref{rhokconditioned}
is given by the analog of the steady state of Eq. \ref{jumptrapstsol}
with the modified trapping times of Eq. \ref{Wdoobjumpeff}
\begin{eqnarray}
{\tilde {\tilde \rho}}^{[\beta]}_y  = \frac{ {\tilde  {\tilde \tau}}^{[\beta]}_y } { \displaystyle \sum_{x=1}^{L}{\tilde  {\tilde \tau}}^{[\beta]}_x}
\label{jumptrapdoobrho}
\end{eqnarray}
Let us now describe special values of $\beta$.


\subsection{ Special value $\beta=0$ } 

For $\beta=0$, Eq. \ref{psijumpeigentrapsol} yields the following value independent of the trapping times
\begin{eqnarray}
    \psi(\beta=0) =1
\label{psiringjumptrapzero}
\end{eqnarray}
This simple value can be understood from the integration of the measure on the first line of Eq. \ref{norma}
where the sums over the positions disappear as a consequence of the one-dimensional directed character of the present directed trap model
\begin{eqnarray}
 \sum_{M=0}^{+\infty}  \int_0^T dt_M \int_0^{t_M} dt_{M-1} ... \int_0^{t_2} dt_{1} 
=\sum_{M=0}^{+\infty} \frac{1}{M!} \prod_{m=1}^M \int_0^T dt_m
= \sum_{M=0}^{+\infty} \frac{T^M}{M!} =e^T
\label{normawt}
\end{eqnarray}
The modified trapping times of Eq. \ref{Wdoobjumpeff} 
\begin{eqnarray}
\frac{ 1 }{{\tilde  {\tilde \tau}}^{[\beta]}_x}  = \psi(\beta=0) =1
  \label{Wdoobjumpeffzero}
\end{eqnarray}
and the corresponding conditioned density of Eq. \ref{jumptrapdoobrho} becomes uniform
\begin{eqnarray}
{\tilde {\tilde \rho}}^{[\beta=0]}_y  = \frac{1}{L}
\label{jumptrapdoobrhozero}
\end{eqnarray}


\subsection{ Limit $\beta \to +\infty$ and the minimal intensive information $I_{min}  $} 

In the limit $\beta \to +\infty$, one expects that $\psi(\beta)$ is negative with the linear behavior of Eq. \ref{zbetainfty}
\begin{eqnarray}
 \psi(\beta)  \opsimeq_{\beta \to + \infty} - \beta I_{min}  
\label{zbetainftyp2}
\end{eqnarray}
The constraint of Eq. \ref{psiringjumptrappos} yields
\begin{eqnarray}
I_{min}  \leq  \frac{ 1 }{\tau_{x}} \ \ \ {\rm for } \ \ \ x=1,2,..,L
\label{psiringchaintrappospinfty}
\end{eqnarray}
while Eq. \ref{psijumpeigentrapsol} becomes
\begin{eqnarray}
0 && \opsimeq_{\beta \to + \infty}   \sum_{x=1}^L \ln \left( \beta \tau^{\beta}_{x}  \left[  \frac{1}{\tau_{x} } - I_{min}\right] \right) 
 \opsimeq_{\beta \to + \infty} L \ln(\beta)  + \beta \sum_{x=1}^L \ln \left(  \tau_{x} \right)
 +
  \sum_{x=1}^L \ln   \left[  \frac{1}{\tau_{x} } - I_{min}\right] 
\label{psijumpeigentrapsolpinfty}
\end{eqnarray}
So the minimum information is determined by the maximal trapping time of the ring occurring at some position $y_{max}$ (Eq. \ref{defymax})
\begin{eqnarray}
 I^{min} =  \frac{ 1 }{\displaystyle \max_{1 \leq y \leq L}\tau_y} =  \frac{ 1 }{\tau_{y_{max}}}
\label{iminexrect}
\end{eqnarray}
in agreement with the direct analysis of Eq. \ref{icase3w},
where the corresponding trajectory with the highest individual trajectory probability of Eq. \ref{maxptraj}
is the trajectory that remains on the site $y_{max}$ (Eq. \ref{maxptrajexw}).
In the Doob generator of the conditioned process,
the modified trapping times of Eq. \ref{Wdoobjumpeff} 
\begin{eqnarray}
{\tilde  {\tilde \tau}}^{[\beta]}_y  \opsimeq_{\beta \to + \infty} 
\frac{ 1 }{ \beta \left( \frac{ 1 }{\tau_y} - I_{min}  \right) }
\opsimeq_{\beta \to + \infty}  
\frac{ 1 }{ \beta \left( \frac{ 1 }{\tau_y} - \frac{ 1 }{\tau_{y_{max}}}  \right) }
 && \opsimeq_{\beta \to + \infty}  1 \ \ \ {\rm if } \ \ y \ne y_{max}
 \nonumber \\
 && \opsimeq_{\beta \to + \infty}  + \infty \ \ \ {\rm if } \ \ y = y_{max} 
  \label{Wdoobjumpeffpinfty}
\end{eqnarray}
vanish at all the sites $y \ne y_{max}$ but diverges for $y=y_{max}$,
so that the corresponding conditioned density of Eq. \ref{jumptrapdoobrho}
is fully localized on the site $y_{max} $
\begin{eqnarray}
{\tilde {\tilde \rho}}^{[\beta]}_y  \opsimeq_{\beta \to + \infty}  \delta_{y,y_{max}}
\label{jumptrapdoobrhoinftypinfty}
\end{eqnarray}
in agreement with the physical interpretation of the localized trajectory of Eq. \ref{maxptrajexw}.


\subsection{ Series expansion in $\beta=1+\epsilon$ up to order $\epsilon^2$ }

For $\beta=1+\epsilon$, the expansion of the logarithm of
 Eq. \ref{psijumpeigentrapsol} up to second order in $\epsilon$ 
\begin{eqnarray}
0 && 
=  \sum_{x=1}^L  \left( \ln \left[    \psi(1+\epsilon)\tau_{x}+ (1+\epsilon)   \right] + \epsilon  \ln(\tau_x) \right)
=  \sum_{x=1}^L  \ln \left[   1 + \epsilon (1+\psi'(1)\tau_x) + \frac{\epsilon^2}{2} \psi''(1) \tau_x   \right]
+ \epsilon   \sum_{x=1}^L \ln(\tau_x)
\nonumber \\
&&  = 
 \sum_{x=1}^L  \left[    \epsilon (1+\psi'(1)\tau_x +\ln(\tau_x) ) + \frac{\epsilon^2}{2} (\psi''(1) \tau_x -(1+\psi'(1)\tau_x)^2 )  \right]
\nonumber \\
&&  = 
\epsilon \left( L +\psi'(1) \sum_{x=1}^L \tau_x + \sum_{x=1}^L \ln(\tau_x) \right) 
+ \frac{\epsilon^2}{2}   \left[ \psi''(1)  \sum_{x=1}^L\tau_x- \sum_{x=1}^L (1+\psi'(1)\tau_x)^2   \right]
\label{psijumpeigentrapsoleps}
\end{eqnarray}

So the order $\epsilon$ allows to recover 
 the Kolmogorov-Sinai entropy of Eq. \ref{hksjumptrap} 
 \begin{eqnarray}
h_{KS} [\tau_{y=1,2,..,L} ]= -\psi'(1)  &&  
=\frac{\displaystyle L+ \sum_{y=1}^{L}       \ln ( \tau_y ) } 
{ \displaystyle \sum_{x=1}^{L} \tau_{x}} 
\label{hksjumptrapprime}
\end{eqnarray}
while the order $\epsilon^2 $ yields the second derivative
\begin{eqnarray}
 \psi''(1) = \frac{ \displaystyle \sum_{y=1}^L (1+\psi'(1)\tau_y)^2  }{ \displaystyle \sum_{x=1}^L\tau_x}  
= 2 \psi'(1) + \frac{ \displaystyle L + [\psi'(1)]^2\sum_{y=1}^L \tau_y^2  }{ \displaystyle \sum_{x=1}^L\tau_x} 
\label{hksjumptrapprime2}
\end{eqnarray}
As a function of the $L$ trapping times $\tau_{y=1,2,..,L}$ of the disordered ring, 
the scaled variance of Eq. \ref{psideri2} reads
\begin{eqnarray}
V_{KS} [\tau_{y=1,2,..,L} ]
=  \frac{ \displaystyle \sum_{y=1}^L (1+\tau_y h_{KS} [\tau_.] )^2  }{ \displaystyle \sum_{x=1}^L\tau_x}  
= -2 h_{KS} [\tau_.] + \frac{ \displaystyle L + h_{KS}^2 [\tau_.] \sum_{y=1}^L \tau_y^2  }{ \displaystyle \sum_{x=1}^L\tau_x} 
\label{varkstrapjump}
\end{eqnarray}
where $h_{KS}[\tau_{y=1,2,..,L} ] $ was given in Eq. \ref{hksjumptrap}.

Let us now analyze its behavior for large $L$ when the probability distribution 
$q(\tau)$ of the trapping times $\tau \in ]1,+\infty[$
is the power-law of Eq. \ref{qtaupower}
depending on the parameter $\mu>0$ : 

(i) in the region $\mu>2$ where the second moment $\overline{\tau^2} $ of the trapping time is finite (Eq. \ref{mom2}),
both the numerator and the denominator of Eq. \ref{varkstrapjump} will follow the law of large numbers
while the Kolmogorov-Sinai entropy converges towards the finite asymptotic value of Eq. \ref{hksjumptraptypfinite}.
As a consequence, the scaled variance of Eq. \ref{varkstrapchain} will then 
converges towards the finite asymptotic value
\begin{eqnarray}
 V_{KS}^{(\infty)}   =   \frac{\displaystyle 
\int_1^{+\infty} d \tau  q(\tau)   \left[ 1+\tau h_{KS}^{(\infty)}  \right]^2 } 
{ \displaystyle \int_1^{+\infty} d \tau \tau q(\tau)}
= -2 h_{KS}^{(\infty)}
+ 
 \frac{\displaystyle 
1+ \left[  h_{KS}^{(\infty)}   \right]^2\int_1^{+\infty} d \tau  q(\tau)     \tau^2   } 
{ \displaystyle \int_1^{+\infty} d \tau \tau q(\tau)}
\ \ \ \ \ {\rm for}  \ \ \ \mu >2
\label{vksjumptraptypfinite} 
\end{eqnarray}

(ii) in the region $1<\mu<2$ where the second moment $\overline{\tau^2} $ of the trapping time is infinite (Eq. \ref{mom2})
while the first moment $\overline{\tau} $ remains finite (Eq. \ref{mom1}),
the only anomalous scaling will come from the sum of the square of the trapping times 
discussed around Eq. \ref{upsilon}.
As a consequence, the scaled variance $V_{KS}$ will not remain finite as in Eq. \ref{vkschaintraptypfinite},
but will diverge with the scaling $L^{\frac{2}{\mu} -1 } $ of exponent $\left(\frac{2}{\mu} -1 \right) \in ]0,1[ $
\begin{eqnarray}
 && V_{KS}^{(L)}  \opsimeq_{L \to + \infty}  L^{\frac{2}{\mu} -1 } \ \vartheta \ 
 \frac{ \left[  h_{KS}^{(\infty)}   \right]^2   } 
{ \displaystyle \int_1^{+\infty} d \tau \tau q(\tau)}
\ \ \ \ \ {\rm for}  \ \ \ 1<\mu <2
\label{varkstrapjumplevy12}
\end{eqnarray}
and will remain distributed over the disordered rings of length $L$ since the variable $\vartheta$ of Eq. \ref{thetalevy1sq}
is distributed with the L\'evy law ${\cal L}_{\frac{\mu}{2}}(\vartheta) $.

(iii) in the region $0<\mu<1$ where both the first moment $\overline{\tau} $
and the second moment $\overline{\tau^2} $ (Eqs \ref {mom1} \ref{mom2}),
while the Kolmogorov-Sinai entropy does not converge anymore towards the finite asymptotic value of Eq. \ref{hksjumptraptypfinite}, one needs to return to the finite-size expression of Eq. \ref{hksjumptrap}
for the Kolmogorov-Sinai entropy
and to re-analyze the leading behavior of Eq. \ref{varkstrapjump}
in terms of the sum $\Sigma_L$ of Eq. \ref{sigmaq} and $\Upsilon_L$ of Eq. \ref{upsilon}
\begin{eqnarray}
 V_{KS}[\tau_{y=1,2,..,L} ]  \opsimeq_{L \to + \infty} 
 \frac{  \displaystyle     \Upsilon_L     }  
{ \displaystyle \Sigma_L^3} L^2  \left( \left[  1+ \int_1^{+\infty} d \tau  q(\tau)       \ln \left( \tau \right) \right] \right)^2
  \oppropto_{L \to + \infty } L^{2-\frac{1}{\mu}  }
  \ \ \ \ {\rm for}  \ \ 0<\mu <1
\label{varkstrapjumplevy01}
\end{eqnarray}
so the scaling in $ L^{2-\frac{1}{\mu}  } $ is different from Eq. \ref{varkstrapjumplevy12},
while the limit distribution would require a more refined analysis of the ratio $ \frac{     \Upsilon_L     }  
{  \Sigma_L^3}  $ involving the two correlated sums of Eq. \ref{sigmaq} and  Eq. \ref{upsilon}.


\subsection{Direct analysis of self-averaging observables in the thermodynamic limit of an infinite ring $L \to +\infty$ } 

As discussed above, the Kolmogorov-Sinai entropy $h_{KS}=-\psi'(1)$ 
is self-averaging in the thermodynamic limit $L \to +\infty$ only for $\mu>1$ (Eq. \ref{hksjumptraptypfinite}),
while the scaled variance $V_{KS}=\psi''(1)$ is self-averaging in the thermodynamic limit $L \to +\infty$ only for $\mu>2$
(Eq. \ref{vksjumptraptypfinite}). Further transitions are expected for the higher cumulants.

However if the trapping time distribution $q(\tau)$ has all its moments finite (in contrast to the power-law form of Eq. \ref{qtaupower}
discussed up to now), then the scaled cumulant generating function $\psi(\beta)$ 
and its derivative will be self-averaging in the thermodynamic limit $L \to +\infty$.
If one rewrites Eq. \ref{psijumpeigentrapsol} via its logarithm and divide by the size $L$ of the ring
\begin{eqnarray}
0 && = \frac{1}{L} \sum_{x=1}^L  \ln\left[ \psi(\beta) \tau^{\beta}_{x}+ \beta \tau^{\beta-1}_{x} \right] 
\label{psijumpeigentrapsollog}
\end{eqnarray}
one obtains that the self-averaging value $\psi_{L=\infty}(\beta)$ in the thermodynamic limit $L \to +\infty$ 
is determined by the equation
\begin{eqnarray}
0 = \int_0^{+\infty} d\tau q(\tau)   \ln \left[ \psi_{\infty}(\beta) \tau^{\beta} + \beta \tau^{\beta-1} \right] 
\label{psithermolimjump}
\end{eqnarray}
However, whenever there are non-self-averaging effects, one should return to the finite-size Eq. \ref{psijumpeigentrapsollog}
to analyze them, as described above for the two first derivatives $\psi'(1) = -h_{KS}$ and $\psi''(1)=V_{KS}$.


\section{ Conclusion }

\label{sec_conclusion}

In this paper, we have revisited the Ruelle thermodynamic formalism for discrete-time Markov chains and for continuous-time Markov Jump processes in the language of the large deviation theory for the intensive information that represents a particularly interesting additive observable of the dynamical trajectories. 
We have described how the generating function of the information can be analyzed 
via the appropriate $\beta$-deformed Markov generator
and the Doob generator of the associated $\beta$-conditioned process. 
In particular, we have stressed that the Kolmogorov-Sinai entropy $h_{KS}$ 
only requires the knowledge of the steady-state $P^*$, 
while the scaled variance $V_{KS}$ of the information requires the knowledge of the Green function $G$.
As examples of applications where all the 
introduced notions can be explicitly evaluated as a function of the parameter $\beta$,
we have focused on the Directed Random Trap Model both in discrete time and in continuous time,
in order to show explicitly that, despite some important technical differences between the two frameworks,
the same conclusions emerge for the physical observables that characterize the glassiness of the dynamics. 
In particular, we have analyzed in detail how the Kolmogorov-Sinai entropy $h_{KS}$ and the scaled variance $V_{KS}$
display anomalous scaling laws with the size $L$ and and non-self-averaging effects in some regions of parameters.,


\appendix

\section{ Reminder on the perturbation theory for an isolated eigenvalue of a non-symmetric matrix }

\label{app_per}

In this Appendix, we consider the expansion of the non-symmetric matrix 
\begin{eqnarray}
M(\epsilon) = M^{(0)} +\epsilon M^{(1)}+ \epsilon^2 M^{(2)} +O(\epsilon^3)
  \label{mseries}
\end{eqnarray}
Th goal is to compute the series expansion of the isolated eigenvalue denoted by the index $0$
\begin{eqnarray}
E_0 (\epsilon) = E_0^{(0)} +\epsilon E_0^{(1)}+ \epsilon^2 E_0^{(2)} +O(\epsilon^3)
  \label{e0series}
\end{eqnarray}
with its corresponding right and left eigenvectors
\begin{eqnarray}
\vert r_0 (\epsilon) \rangle && = \vert r_0^{(0)}  \rangle +\epsilon \vert r_0^{(1)}\rangle + \epsilon^2 \vert r_0^{(2)} \rangle +O(\epsilon^3)
\nonumber \\
\langle  l_0 (\epsilon) \vert && = \langle  l_0^{(0)}  \vert  +\epsilon \langle  l_0^{(1)}  \vert + \epsilon^2 \langle  l_0^{(2)}  \vert  +O(\epsilon^3)
  \label{rl0series}
\end{eqnarray}

\subsection { Eigenvalue equations and normalization of the eigenvectors  }

One writes the series expansion of the eigenvalue equation for the right eigenvector
\begin{eqnarray}
0  = \left( M(\epsilon) - E_0 (\epsilon)  \right) \vert r_0 (\epsilon) \rangle 
&& = 
\left( M^{(0)} - E_0^{(0)}  \right) \vert r_0^{(0)} \rangle
+ \epsilon
\left[\left( M^{(0)} - E_0^{(0)}  \right) \vert r_0^{(1)} \rangle + \left( M^{(1)} - E_0^{(1)}  \right) \vert r_0^{(0)} \rangle
\right]
\nonumber \\
&&
\! \! \! + \epsilon^2
\left[\left( M^{(0)} - E_0^{(0)}  \right) \vert r_0^{(2)} \rangle + \left( M^{(1)} - E_0^{(1)}  \right) \vert r_0^{(1)} \rangle
+\left( M^{(2)} - E_0^{(2)}  \right) \vert r_0^{(0)} \rangle
\right]+O(\epsilon^3)
  \label{powerright}
\end{eqnarray}
and for the left eigenvector
\begin{eqnarray}
0  = \langle  l_0 (\epsilon)\vert\left( M(\epsilon) - E_0 (\epsilon)  \right) 
&& = 
\langle  l_0^{(0)} \vert \left( M^{(0)} - E_0^{(0)}  \right) 
+ \epsilon
\left[ \langle  l_0^{(1)} \vert \left( M^{(0)} - E_0^{(0)}  \right)  +\langle  l_0^{(0)} \vert  \left( M^{(1)} - E_0^{(1)}  \right) 
\right]
\nonumber \\
&&
+ \epsilon^2
\left[\langle  l_0^{(2)} \vert\left( M^{(0)} - E_0^{(0)}  \right) 
 + \langle  l_0^{(1)} \vert\left( M^{(1)} - E_0^{(1)}  \right) 
+\langle  l_0^{(0)} \vert\left( M^{(2)} - E_0^{(2)}  \right) 
\right]+O(\epsilon^3)
  \label{powerleft}
\end{eqnarray}
as well as the series expansion of the normalization 
\begin{eqnarray}
1= \langle  l_0 (\epsilon)\vert r_0 (\epsilon) \rangle && = 
\langle  l_0^{(0)}  \vert r_0^{(0)}  \rangle
+ \epsilon \left( \langle  l_0^{(0)}  \vert r_0^{(1)}  \rangle  + \langle  l_0^{(1)}  \vert r_0^{(0)}  \rangle\right) 
+ \epsilon^2 \left( \langle  l_0^{(0)}  \vert r_0^{(2)}  \rangle  + \langle  l_0^{(1)}  \vert r_0^{(1)}  \rangle
+\langle  l_0^{(0)}  \vert r_0^{(2)}  \rangle \right)
+O(\epsilon^3)
  \label{powernorma}
\end{eqnarray}

For $\epsilon=0$, one assumes that one knows the unperturbed eigenvalue $E_0^{(0)} $ of the unperturbed matrix $M^{(0)}$
together with its right and left eigenvectors properly normalized
\begin{eqnarray}
0  && =  \left( M^{(0)} - E_0^{(0)}  \right) \vert r_0^{(0)} \rangle
\nonumber \\
0  && = \langle  l_0^{(0)} \vert \left( M^{(0)} - E_0^{(0)}  \right) 
\nonumber \\
1 &&  = \langle  l_0^{(0)}  \vert r_0^{(0)}  \rangle
  \label{powerzero}
\end{eqnarray}

\subsection { First-order perturbation }

At order $\epsilon$, the standard choice that respects the normalization of Eq \ref{powernorma}
is given by the orthogonality conditions for the first-order corrections
of the eigenvectors with respect to the unperturbed eigenvectors
\begin{eqnarray}
0 && =  \langle  l_0^{(0)}  \vert r_0^{(1)}  \rangle  
\nonumber \\
0 && =  \langle  l_0^{(1)}  \vert r_0^{(0)}  \rangle
  \label{powernorma1}
\end{eqnarray}
Then the eigenvalue Eq. \ref{powerright} for the right eigenvector at order $\epsilon$
\begin{eqnarray}
0 = \left( M^{(0)} - E_0^{(0)}  \right) \vert r_0^{(1)} \rangle + \left( M^{(1)} - E_0^{(1)}  \right) \vert r_0^{(0)} \rangle
  \label{powerright1}
\end{eqnarray}
can be projected onto the unperturbed left eigenvector $\langle  l_0^{(0)} \vert $
to obtain the first-order correction of the eigenvalue
\begin{eqnarray}
E_0^{(1)} = \langle  l_0^{(0)} \vert M^{(1)}  \vert r_0^{(0)} \rangle
  \label{energy1}
\end{eqnarray}
Equivalently, the eigenvalue Eq. \ref{powerleft} for the right eigenvector at order $\epsilon$
\begin{eqnarray}
0  =  \langle  l_0^{(1)} \vert \left( M^{(0)} - E_0^{(0)}  \right)  +\langle  l_0^{(0)} \vert  \left( M^{(1)} - E_0^{(1)}  \right) 
  \label{powerleft1}
\end{eqnarray}
can be projected onto the unperturbed right eigenvector $\vert r_0^{(0)} \rangle$ to obtain again Eq. \ref{energy1}.

To obtain the first-order corrections of the eigenvectors, 
one needs to introduce the inverse of the operator $( E_0^{(0)} - M^{(0)} )$
within the subspace orthogonal to the subspace $( \vert r_0^{(0)} \rangle \langle  l_0^{(0))} \vert )  $ of the unperturbed eigenvectors
\begin{eqnarray}
G^{(0)} \equiv \left( \mathbb{1} - \vert r_0^{(0)} \rangle \langle  l_0^{(0)} \vert \right)
\frac{1 }{  E_0^{(0)} \mathbb{1} - M^{(0)}  } 
\left( \mathbb{1} - \vert r_0^{(0)} \rangle \langle  l_0^{(0)} \vert \right)
  \label{green}
\end{eqnarray}
The application of this Green function $G^{(0)}$ on the left of Eq. \ref{powerright1}
and on the right of Eq. \ref{powerleft1}
yields that the first-order corrections for the eigenvectors read using Eq. \ref{powernorma1}
\begin{eqnarray}
 \vert r_0^{(1)} \rangle && = G^{(0)}  M^{(1)}  \vert r_0^{(0)} \rangle
 \nonumber \\
   \langle  l_0^{(1)} \vert  && = \langle  l_0^{(0)} \vert  M^{(1)}  G^{(0)}
  \label{resrl11}
\end{eqnarray}
If the Green function $G^{(0)}$ from Eq. \ref{green} is computed in the basis of the eigenvectors of 
the unperturbed matrix $M^{(0)}$
\begin{eqnarray}
G^{(0)} = \sum_{k \ne 0} 
\frac{\vert r_k^{(0)} \rangle \langle  l_k^{(0)} \vert  }{  E_0^{(0)} - E_k^{(0)}  } 
  \label{greenspectral}
\end{eqnarray}
one recovers the analog of the familiar formulas from quantum mechanics perturbation theory.
However, if one does not know the full spectrum of the unperturbed matrix $M^{(0)}$
or if one does not wish to compute it,
the Green function $G^{(0)}$ of Eq. \ref{green} can be computed directly by solving the matrix equations
\begin{eqnarray}
\left( E_0^{(0)}\mathbb{1} - M^{(0)}  \right) G^{(0)} && =  \mathbb{1} - \vert r_0^{(0)} \rangle \langle  l_0^{(0)} \vert 
\nonumber \\
G^{(0)} \left( E_0^{(0)}\mathbb{1} - M^{(0)}  \right) && =  \mathbb{1} - \vert r_0^{(0)} \rangle \langle  l_0^{(0)} \vert 
  \label{eqgreen}
\end{eqnarray}
with the orthogonality conditions
\begin{eqnarray}
\langle  l_0^{(0)} \vert  G^{(0)} && =  0
\nonumber \\
G^{(0)}  \vert r_0^{(0)} \rangle && =  0
  \label{eqgreenortho}
\end{eqnarray}

\subsection { Second-order perturbation }

The eigenvalue Eq. \ref{powerright} for the right eigenvector at order $\epsilon^2$
\begin{eqnarray}
0= \left( M^{(0)} - E_0^{(0)}  \right) \vert r_0^{(2)} \rangle + \left( M^{(1)} - E_0^{(1)}  \right) \vert r_0^{(1)} \rangle
+\left( M^{(2)} - E_0^{(2)}  \right) \vert r_0^{(0)} \rangle
  \label{powerright2}
\end{eqnarray}
can be projected onto the unperturbed left eigenvector $\langle  l_0^{(0)} \vert $
to obtain the second-order correction of the eigenvalue
\begin{eqnarray}
E_0^{(2)} = \langle  l_0^{(0)}  \vert  M^{(1)}  \vert r_0^{(1)} \rangle+\langle  l_0^{(0)}  \vert M^{(2)}  \vert r_0^{(0)} \rangle
  \label{energy2r}
\end{eqnarray}
Similarly, the projection of the eigenvalue
Eq \ref{powerleft} for the left eigenvector at order $\epsilon^2$
\begin{eqnarray}
0= \langle  l_0^{(2)} \vert\left( M^{(0)} - E_0^{(0)}  \right) 
 + \langle  l_0^{(1)} \vert\left( M^{(1)} - E_0^{(1)}  \right) 
+\langle  l_0^{(0)} \vert\left( M^{(2)} - E_0^{(2)}  \right) 
  \label{powerleft2}
\end{eqnarray}
onto the unperturbed right eigenvector $\vert r_0^{(0)} \rangle $
yields
\begin{eqnarray}
E_0^{(2)} =  \langle  l_0^{(1)} \vert M^{(1)}  \vert r_0^{(0)} \rangle
+\langle  l_0^{(0)} \vert M^{(2)}  \vert r_0^{(0)} \rangle
  \label{energy2l}
\end{eqnarray}

Using the firs-order corrections of the eigenvectors of Eq. \ref{resrl11},
one obtains that Eqs \ref{energy2l}
and \ref{energy2l}
give the same final result 
\begin{eqnarray}
E_0^{(2)} =  \langle  l_0^{(0)} \vert  M^{(1)}  G^{(0)} M^{(1)}  \vert r_0^{(0)} \rangle
+\langle  l_0^{(0)} \vert M^{(2)}  \vert r_0^{(0)} \rangle
  \label{energy2}
\end{eqnarray}
as it should for consistency.
If one uses the decomposition of Eq. \ref{green} for the Green function $G^{(0)}$,
one recovers the analog of the familiar formula from quantum mechanics perturbation theory,
but $G^{(0)}$ can also be computed directly from Eqs \ref{eqgreen} and \ref{eqgreenortho}.


\section{ Reminder on L\'evy sums }

\label{app_levy}

Let us assume that the probability distribution $q(\tau)$ of the trapping time $\tau \in ]1,+\infty[$
is the power-law depending on the parameter $\mu>0$
\begin{eqnarray}
q(\tau) =   \frac{ \mu}{\tau^{1+\mu} }
\label{qtaupower}
\end{eqnarray}
The non-integer moments of order $k$ are finite only for $k<\mu$
\begin{eqnarray}
\overline{\tau^k} = \int_1^{+\infty} d \tau \tau^k q(\tau) =   \frac{ \mu}{ \mu-k } \ \ \ \ \ {\rm for}  \ \ \ k<\mu
\label{mom}
\end{eqnarray}
and diverge for $k \geq \mu$.
In particular, the first moment $k=1$ is finite only for $\mu>1$
\begin{eqnarray}
\overline{\tau} = \int_1^{+\infty} d \tau \tau q(\tau) =   \frac{ \mu}{ \mu-1 } \ \ \ \ \ {\rm for}  \ \ \ \mu >1
\label{mom1}
\end{eqnarray}
while the second moment $k=2$ is finite only for $\mu>2$
\begin{eqnarray}
\overline{\tau^2} = \int_1^{+\infty} d \tau \tau^2 q(\tau) =   \frac{ \mu}{ \mu-2 } \ \ \ \ \ {\rm for}  \ \ \ \mu>2
\label{mom2}
\end{eqnarray}
as the variance
\begin{eqnarray}
\sigma^2 \equiv  \overline{\tau^2} - (\overline{\tau})^2 
=  \frac{ \mu}{ \mu-2 } - 
 \left(  \frac{ \mu}{ \mu-1 }\right)^2 
 \ \ \ \ \ {\rm for}  \ \ \ \mu>2
\label{vartau}
\end{eqnarray}

As a consequence, the sum of $L$ independent trapping times
\begin{eqnarray}
\Sigma_L \equiv \sum_{x=1}^L \tau_x
\label{sigmaq}
\end{eqnarray}
will have a finite averaged value only for $\mu>1$
\begin{eqnarray}
\overline{\Sigma_L} = L  \overline{\tau} = L  \frac{ \mu}{ \mu-1 } \ \ \ \ \ {\rm for}  \ \ \ \mu >1
\label{smom1}
\end{eqnarray}
and a finite variance only for $\mu>2$
\begin{eqnarray}
\overline{\Sigma_L^2} - ( \overline{\Sigma_L} )^2 
= L \left( \overline{\tau^2} - (\overline{\tau})^2 \right)  =   L \sigma^2
 \ \ \ \ \ {\rm for}  \ \ \ \mu>2
\label{smom2}
\end{eqnarray}
So the Central Limit Theorem will be valid only in the region $\mu>2$, where the 
the appropriate rescaled variable 
\begin{eqnarray}
\theta \equiv \frac{\Sigma_L- L \overline{\tau}}{ \sqrt{L} \sigma }
\label{thetaCLT}
\end{eqnarray}
will be Gaussian distributed.

For $0<\mu<2$, the sum $\Sigma_L$ can be analyzed instead in terms
of L\'evy stables laws. Note that L\'evy sums have been analyzed in the context of various disordered systems
\cite{jpbreview,Der,Der_Fly,us_critiweights,us_multifandersonlevy,c_rgsglevy,c_levymatrix}.


\subsection{ L\'evy sum $ \Sigma_L$  for  $0<\mu <1$ }

For $0<\mu<1$, the Laplace transform of the distribution of Eq. \ref{qtaupower}
presents the characteristic singularity in $p^{\mu}$ in the Laplace variable $p$ near the origin $p \to 0$
\begin{eqnarray}
\overline{e^{-p \tau} } && = \int_1^{+\infty} d \tau  q(\tau) e^{-p \tau}
= 1 -  \int_1^{+\infty} d \tau    \frac{ \mu}{\tau^{1+\mu} } (1- e^{-p \tau} )
= 1 -  p^{\mu} \int_p^{+\infty} dt    \frac{ \mu}{t^{1+\mu} } (1- e^{- t} )
\nonumber \\
&&  \opsimeq_{p \to 0} 1- p^{\mu} \int_0^{+\infty} dt    \frac{ \mu}{t^{1+\mu} } (1- e^{- t} ) +... 
  \opsimeq_{p \to 0} e^{- p^{\mu} \int_0^{+\infty} dt    \frac{ \mu}{t^{1+\mu} } (1- e^{- t} ) +... }
\label{lawzqlaplace}
\end{eqnarray}
So the generating function of the sum of Eq. \ref{sigmaq}
 will display the same singularity
\begin{eqnarray}
\overline{e^{-p \Sigma_L } }  && = \left(\overline{e^{-p \tau} } \right)^L
  \opsimeq_{p \to 0} e^{- L p^{\mu} \int_0^{+\infty} dt    \frac{ \mu}{t^{1+\mu} } (1- e^{- t} ) +... } 
  = e^{- \left( L^{\frac{1}{\mu} } p\right)^{\mu} \int_0^{+\infty} dt    \frac{ \mu}{t^{1+\mu} } (1- e^{- t} ) +... }
\label{sigmaqlaplace}
\end{eqnarray}
This means that the sum $ \Sigma_L $ will grow as $ L^{\frac{1}{\mu} } $,
i.e. more rapidly than linearly in $L$,
and that the appropriate rescaled variable 
\begin{eqnarray}
\theta \equiv \frac{\Sigma_L}{ L^{\frac{1}{\mu} } }
\label{thetalevy1}
\end{eqnarray}
will be distributed with the L\'evy law ${\cal L}_{\mu}(\theta)$ of index $\mu \in ]0,1[$ determined by the Laplace transform
\begin{eqnarray}
 \int_0^{+\infty} d \theta {\cal L}_{\mu}(\theta) e^{-p \theta}  = e^{-  p^{\mu}  \int_0^{+\infty} dt    \frac{ \mu}{t^{1+\mu} } (1- e^{- t} )}
\label{levy1laplace}
\end{eqnarray}


\subsection{ L\'evy sum $ \Sigma_L$ for $1<\mu <2$ }

In the region $1<\mu <2$, the averaged value $ \overline{\tau} $ of Eq. \ref{mom1} exists.
In the Laplace transform of the distribution of Eq. \ref{qtaupower},
the singularity in $p^{\mu}$ will thus appear after the regular term in $p$
\begin{eqnarray}
\overline{e^{-p \tau} } && = \int_1^{+\infty} d \tau  q(\tau) e^{-p \tau}
= 1 -  p  \overline{\tau} 
- \int_1^{+\infty} d \tau    \frac{ \mu}{\tau^{1+\mu} } (1  -  p  \tau  - e^{-p \tau} )
= 1 -  p  \overline{\tau} -  p^{\mu} \int_p^{+\infty} dt    \frac{ \mu}{t^{1+\mu} } (1 -  t - e^{- t} )
\nonumber \\
&&  \opsimeq_{p \to 0} 1 -  p  \overline{\tau} -  p^{\mu} \int_0^{+\infty} dt    \frac{ \mu}{t^{1+\mu} } (1 -  t - e^{- t} ) +... 
  \opsimeq_{p \to 0} e^{-  p  \overline{\tau} -  p^{\mu} \int_0^{+\infty} dt    \frac{ \mu}{t^{1+\mu} } (1 -  t - e^{- t} ) +... }
\label{lawzqlaplace2}
\end{eqnarray}
So the generating function of the sum of Eq. \ref{sigmaq}
 will display the same singularity
\begin{eqnarray}
\overline{e^{-p \Sigma_L } }  && = \left(\overline{e^{-p \tau} } \right)^L
  \opsimeq_{p \to 0} e^{-  L p  \overline{\tau} -  L p^{\mu} \int_0^{+\infty} dt    \frac{ \mu}{t^{1+\mu} } (1 -  t - e^{- t} ) +... }
\label{sigmaqlaplace2}
\end{eqnarray}
This means that the difference $ (\Sigma_L - L \overline{\tau}) $ will scale as $ L^{\frac{1}{\mu} } $,
i.e. will be bigger than the Central-Limit fluctuations of order $L^{\frac{1}{2} } $ of Eq. \ref{thetaCLT}.
The appropriate rescaled variable 
\begin{eqnarray}
\theta \equiv \frac{\Sigma_L- L \overline{\tau}}{ L^{\frac{1}{\mu} } }
\label{thetalevy2}
\end{eqnarray}
will be distributed with the L\'evy law ${\cal L}_{\mu}(\theta)$ of index $\mu \in ]1,2[$ determined by the Laplace transform
\begin{eqnarray}
 \int_0^{+\infty} d \theta {\cal L}_{\mu}(\theta) e^{-p \theta}  = 
  e^{-  L p^{\mu} \int_0^{+\infty} dt    \frac{ \mu}{t^{1+\mu} } (1 -  t - e^{- t} ) }
 \label{levy2laplace}
\end{eqnarray}


\subsection{ Translation for the sum $\Upsilon_L $ of the squares of the trapping times  }

In the text, we will also need to analyze the sum of the squares of the trapping times 
\begin{eqnarray}
\Upsilon_L \equiv \sum_{x=1}^L \tau^2_x
\label{upsilon}
\end{eqnarray}
Since the variable $u=\tau^2 \in ]1,+\infty[$ is distributed with the probability
\begin{eqnarray}
Q(u) =   \frac{ \frac{\mu}{2} }{u^{1+\frac{\mu}{2} } }
\label{qupower}
\end{eqnarray}
that corresponds to the power-law form of Eq. \ref{qtaupower} but with the modified parameter $\mu'= \frac{\mu}{2}$,
the previous discussion can be directly translated as follows :

(i) the Central-Limit theorem will be valid for the sum of Eq. \ref{upsilon}
only in the region $\mu'>2$ corresponding to $\mu>4$

(ii) in the region $1<\mu'<2$ corresponding to $2<\mu<4$,
the appropriate rescaled variable 
\begin{eqnarray}
\vartheta \equiv \frac{\Upsilon_L- L \overline{\tau^2}}{ L^{\frac{1}{\mu'} } } =  \frac{\Upsilon_L- L \overline{\tau^2}}{ L^{\frac{2}{\mu} } }
\label{thetalevy2sq}
\end{eqnarray}
will be distributed with the L\'evy law ${\cal L}_{\mu'}(\vartheta)={\cal L}_{\frac{\mu}{2}}(\vartheta) $ of index $\mu' \in ]1,2[$

(iii) in the region $0<\mu'<1$ corresponding to $0<\mu<2$,
the appropriate rescaled variable 
\begin{eqnarray}
\vartheta \equiv \frac{\Upsilon_L}{ L^{\frac{1}{\mu'} } } =\frac{\Upsilon_L}{ L^{\frac{2}{\mu} } }
\label{thetalevy1sq}
\end{eqnarray}
will be distributed with the L\'evy law ${\cal L}_{\mu'}(\vartheta)={\cal L}_{\frac{\mu}{2}}(\vartheta) $ of index $\mu' \in ]0,1[$.


\end{document}